\documentclass[12pt,a4paper]{article}
\DeclareMathAlphabet{\scr}{U}{rsfs}{m}{n}
\usepackage{supertabular}
\usepackage{amsmath}
\usepackage{amssymb}
\usepackage{wasysym}
\usepackage{bbold}
\usepackage[sans]{dsfont}
\usepackage{graphicx}
\usepackage{epsfig}
\usepackage[dvips]{color}

\newcommand{\cleqn}{\setcounter{equation}{0}}
\newcommand{\newc}{\newcommand}
\newc{\eps}{\epsilon}
\newc{\Lam}{\Lambda}
\newc{\ra}{\rightarrow}
\newc{\wtilde}{\widetilde}
\newc{\ie}{\textit{i.e.}}
\newc{\cf}{\textit{cf.}}
\newc{\beq}{\begin{equation}}
\newc{\eeq}{\end{equation}}
\newc{\beqn}{\begin{eqnarray}}
\newc{\eeqn}{\end{eqnarray}}
\newc{\PLB}{\emph{Phys.Lett.}{\bf{B}}}
\newc{\NPB}{\emph{Nucl.Phys.}{\bf{B}}}
\newc{\mcal}{\mathcal}
\newc{\nonum}{\nonumber}
\newc{\lsim}{\apprle}
\newc{\gsim}{\apprge}
\newc{\gev}{{\,{\rm GeV}}}
\newc{\lam}{\lambda}
\newc{\kap}{\kappa}
\newc{\ups}{\upsilon}
\newc{\bsym}{\boldsymbol}
\newc{\mc}{\mathcal}
\definecolor{grey}{rgb}{0.8,0.8,0.7}
\definecolor{Red}{cmyk}{0,1,1,0}
\definecolor{luhn}{rgb}{1,0,0.5}
\newc{\luhn}{\color{luhn}}
\newc{\maZ}{\boldsymbol{Z}}
\newc{\Mrm}{\mathrm}
\newc{\Gsm}{G_{\mathrm{SM}}}
\begin{document}
\title{\textbf{What is the Discrete Gauge Symmetry \\ of the MSSM?}}
\date{}
\author{Herbi K. Dreiner$^{1}$,\footnote{
    E-Mail: {\tt dreiner@th.physik.uni-bonn.de}} 
    ~~Christoph Luhn$^{1}$,\footnote{
    E-Mail: {\tt luhn@th.physik.uni-bonn.de}} 
    ~~Marc Thormeier$^{2}$\footnote{
    E-Mail: {\tt marc.thormeier@cea.fr}}\\[2ex]
\small\it $^1$Physikalisches Institut der 
Universit\"at Bonn,   Nu\ss{}allee 12, 53115 Bonn, Germany\\
\small\it $^2$Service de Physique Th\'eorique, 
CEA-Saclay,   91191 Gif-sur-Yvette Cedex, France}
\maketitle

\begin{abstract}
  We systematically study the extension of the Supersymmetric Standard
  Model (SSM) by an anomaly-free discrete gauge symmetry $\maZ_N$. We
  extend the work of Ib\'a\~nez and Ross with $N\!=\!2,3$ to arbitrary
  values of $N$. As new fundamental symmetries, we find four $\maZ_6$,
  nine $\maZ_9$ and nine $\maZ_{18}$. We then place three
  phenomenological demands upon the low-energy effective SSM: {\it
  (i)} the presence of the $\mu$-term in the superpotential, {\it
  (ii)} baryon-number conservation up to dimension-five operators, and
  {\it (iii)} the presence of the see-saw neutrino mass term
  $LH_uLH_u$. We are then left with only two anomaly-free discrete
  gauge symmetries: baryon-triality, $\bsym{B}_3$, and a new $\maZ_6$,
  which we call proton-hexality, $\bsym{P}_6$. Unlike $\bsym{B}_3$,
  $\bsym{P}_6$ prohibits the dimension-four lepton-number violating
  operators. This we propose as {\it the} discrete gauge symmetry of
  the Minimal SSM, instead of $R$-parity.
\end{abstract}

\section{Introduction}
\label{intro}
\cleqn
\noindent
The action of the Standard Model (SM)
\cite{Weinberg:1967tq,Glashow:1961tr} is invariant under Poincar\'e
transformations, as well as the gauge group $\Gsm=SU(3)_C \times
SU(2)_W\times U(1)_Y$. When allowing only renormalisable interactions,
baryon- and lepton-number are (accidental) global symmetries of the
SM.\footnote{\label{firstfootnote}When taking into account the
  sphaleron interactions \cite{Manton:1983nd}, only $\frac{1}{3}B-L_
  i$, and $L_i-L_j$ are conserved in the SM. For the effect of
  sphaleron interactions in supersymmetry see for example
  Refs.~\cite{Harvey:1990qw,Dreiner:1992vm,Laine:1999wv}.} However,
when considering the SM as a low-energy effective theory, $\Gsm$
allows for non-renormalisable interactions, which can violate lepton-
and baryon-number. The leading dimension-six operators are suppressed
by two powers of an unknown mass scale $M$, which is unproblematic for
proton decay if $M\gsim 10^{16}\gev$, see however 
  \cite{Dorsner:2004xa,Nath:2006ut}.

Enlarging the Poincar\'e group, the action of the Supersymmetric SM
(SSM) is invariant under supersymmetry, as well as $\Gsm$
\cite{Nilles:1983ge,Martin:1997ns}. The \textit{renormalisable}
superpotential of the SSM is given by
\cite{Sakai:1981pk,Weinberg:1981wj,Dreiner:1997uz,Dreiner:1991pe}
\beqn
W&=& h^E_{ij}\; L_i H_d {\bar E}_j + h^D_{ij}\; Q_i H_d {\bar D}_j +
h^U_{ij}\; Q_i H_u {\bar U}_j + \mu\,H_d H_u \nonum
\\ 
&+&\!\!\lam_{ijk}\;L_i  L_j {\bar E}_k+ \lam^\prime_{ijk}\, 
L_i Q_j {\bar D}_k + \lam''_{ijk}\,{\bar U}_i{\bar D}_j{\bar D}_k
+ \kap_i\, L_i H_u\,,
\label{superpot}
\eeqn
where we employ the notation of Ref.~\cite{Allanach:2003eb}, and
$SU(3)_C$ and $SU(2)_W$ indices are suppressed. The fifth, sixth and
eighth terms violate lepton-number, and the seventh term violates
baryon-number.  Thus in the SSM, lepton- and baryon-number are
violated by {renormalisable} dimension-four interactions. In
particular, $LQ{\bar D}$ and ${\bar U}{\bar D}{\bar D}$ together lead
to rapid proton decay. The lower experimental bound on the proton
lifetime \cite{Shiozawa:1998si,Hayato:1999az} results in the very
stringent bounds \cite{Goity:1994dq,Dreiner:1997uz,Perez:2004th}
\beq 
\lam'_{i1j}\cdot\lam''_{11j} < 2\cdot 10^{-27} \left(\frac{M_{{
\tilde d}_j}}{100\gev}\right)^2\,,\quad i=1,2\,,\; j\not=1\,, 
\label{bound}
\eeq
and the SSM must be considered incomplete. In order to obtain a
natural and viable supersymmetric model, we must extend $\Gsm$, such
that at least one of the operators $LQ{\bar D} $ or ${\bar U}{\bar
  D}{\bar D}$ is forbidden.\footnote{For an extensive set of bounds on
  the products of these operators see
  Refs.~\cite{Smirnov:1996bg,Allanach:1999ic}.}

The Minimal SSM (MSSM) is conventionally taken as the renormalisable
SSM with the superpotential, Eq.~(\ref{superpot}), additionally
constrained by the discrete symmetry $R$-parity, $\bsym{R}_p=(-\bsym{1
})^{2S+3B+L}$ \cite{Farrar:1978xj}, which acts on the components of
the superfields. Here $S$ is spin, $B$ baryon-number and $L$
lepton-number. Hence the superpotential of the renormalisable MSSM is
given solely by the first line of Eq.~(\ref{superpot}), and baryon-
and lepton-number are conserved. Matter-parity ($\bsym{M}_p$)
\cite{Bento:1987mu}, acts on the superfields and leads to the same
superpotential as $\bsym{R}_p$. Our working definition of the MSSM
shall be the SSM constrained by $\bsym{M}_p$. We return to this in
Sect.~\ref{physics}. Another possibility is to extend $\Gsm$ by
baryon-triality\footnote{This was originally introduced as
baryon-parity in \cite{Ibanez:1991hv,Ibanez:1991pr}; however, it is
more appropriately called baryon-triality
\cite{Martin:1997ns,Grossman:1998py}.}  ($\bsym{B}_3$)
\cite{Ibanez:1991hv,Ibanez:1991pr}, leading to the $R$-parity violating
MSSM \cite{Allanach:2003eb}.

However, due to the unification of the $\Gsm$ gauge coupling constants
in supersymmetry
\cite{Langacker:1991an,Ellis:1990wk,Amaldi:1991cn,Allanach:1999mh},
and also the automatic inclusion of gravity in local supersymmetry
\cite{Freedman:1976xh,Deser:1976eh}, we expect the SSM, and also the
MSSM, to be low-energy effective theories, embedded in a more complete
theory formulated at the scale of Grand Unified Theories ($M_{
  \mathrm{GUT}}\sim 10^{16}\gev$) \cite{Georgi:1974sy}, or above.
Within the SSM, we must therefore take into account the possible
non-renormalisable operators, which are consistent with $\Gsm$, within
the MSSM, those which are also consistent with $\bsym{M}_p$.~In
particular, we are here interested in the dimension-five baryon-
and/or lepton-number violating interactions. In Eq.~(\ref{dim5}), we
list the complete set for the SSM
\cite{Sakai:1981pk,Weinberg:1981wj,Allanach:2003eb,Ibanez:1991pr}; a
subset is also present in the MSSM. Even if suppressed by the gravitational
scale $M_{\mathrm{grav}} = 2.4 \times 10^{18}~\mathrm{GeV}$, these operators 
are potentially dangerous, depending on their
flavour structure \cite{Sakai:1981pk,Weinberg:1981wj,Harnik:2004yp}.
Thus, even though $\bsym{M}_p$ provides the SSM with an excellent
candidate for cold dark matter it has a serious problem with
baryon-number violation.  When considering the (high-energy) symmetry
extension of the SSM, we take into account the effects on the
dimension-four and the dimension-five operators.

It is the purpose of this paper to systematically investigate discrete
$\maZ_N$~symmetry extensions of $\Gsm$ without invoking the existence
of new light particles. Since a global discrete symmetry is typically
violated by quantum gravity effects \cite{Krauss:1988zc}, we focus on
an Abelian discrete \textit{gauge} symmetry (DGS): it is a discrete
remnant~of a spontaneously broken $U(1)$ gauge symmetry
\cite{Krauss:1988zc,Preskill:1990bm}. For an explicit Lagrangian see,
\textit{e.g.}, Ref.~\cite{Banks:1989ag}.  Assuming the original gauge
theory to be anomaly-free, Ib\'a\~nez and Ross (IR) determined the
constraints on the remnant low-energy and family-independent DGSs
\cite{Ibanez:1991hv,Ibanez:1991pr}. They classified all $\maZ_N$ DGSs
for $N=2,3$ according to their action on the baryon- and lepton-number
violating operators and then determined which are discrete gauge
anomaly-free (see the end of Sect.~\ref{linear}). They found only two
such anomaly-free DGSs which prohibited the dimension-four
baryon-number violating operators and allowed the $H_dH_u$ term:
matter-parity ($R_2$ in their notation) and baryon-triality,
$\bsym{B}_3 $. The latter has the advantage of also prohibiting the
dangerous dimension-five operators.

In this paper, we extend the work of IR to $\maZ_N$~symmetries with
arbitrary values of $N$. We first determine all family-independent
anomaly-free DGSs consistent with the first three terms in
Eq.~(\ref{superpot}) (Sects.~\ref{linear}-\ref{cubic}). From the
low-energy point of view, where heavy and possibly $\maZ_N$ charged
particles do not play a r\^ole, this infinite number of anomaly-free
DGSs can be rescaled to an equivalent finite set, which we denote as
fundamental (Sect.~\ref{rescaling}). We are left with four $\maZ_6$,
nine $\maZ_9$, and nine $\maZ_{18} $ new symmetries, beyond the five
$\maZ_{2,3}$~symmetries of IR. Together these twenty-seven fundamental
DGSs comprise a complete set. This is one of the main results of this
paper. Next, we investigate their effect on the baryon- and lepton-number 
violating operators (Sect.~\ref{physics}). There is only one
DGS which simultaneously allows the $H_dH_u$ term, prohibits all
dimension-four baryon- and lepton-number violating operators,
prohibits the dimension-five baryon-number violating operators and
{\it allows} the dimension-five Majorana neutrino mass term $LH_uLH_
u$. This is one of the $\maZ_6$ symmetries, $R_6^5L_6^2$, in the
notation of IR.  We shall denote it proton-hexality, $\bsym{P}_6$.
\emph{This we propose as the DGS of the MSSM.} Every $\maZ_6$ is
isomorphic to a direct product of a $\maZ_2$ and a $\maZ_3$
\cite{Kurzweil:1977}, so it is not too surprising that $\bsym{P}_6$ is
isomorphic to the direct product of $\bsym{M}_p$ and $\bsym{B}_3$.  We
then investigate the necessity of heavy fermions in theories with
anomaly-free DGSs (Sect.~\ref{heavy}), leading to a different
conclusion than Ref.~\cite{Ibanez:1992ji}.

In Sects.~\ref{linear}-\ref{heavy} we take a bottom-up approach in
determining the discrete symmetry. At the CERN LHC, we will hopefully
discover supersymmetric fields and their interactions. Through the
measured and thus allowed interactions we can infer the discrete
symmetry. From this point-of-view, two discrete symmetries are
equivalent, if they result in the same low-energy interactions. In
Sect.~\ref{top-down}, we instead investigate the top-down perspective,
focussing on the distinct gauge theories leading to low-energy
equivalent DGSs.  For demonstrational purposes we finally present a
gauged $U(1)$ model, which, after spontaneous symmetry breaking, leads
to an effective SSM with proton-hexality (Sect.~\ref{gauged}).

We briefly comment on some related work in the literature.  Throughout
we restrict ourselves to family-independent DGSs. For examples of
family-dependent DGSs see
Refs.~\cite{Ibanez:1991pr,Kapetanakis:1992jj}.  We shall, however, in
general, allow for the original gauge symmetry to be family-dependent.
We do not consider discrete $R$-symmetries.  For an anomaly-free
gauged $U(1)$ $R$-symmetry in a local supersymmetric theory see
Refs.~\cite{Chamseddine:1995gb,Castano:1995ci,Kurosawa:2001iq}. This 
could be broken to a discrete $R$-symmetry. Since $R$-parity is inserted 
\textit{ad hoc} in the SSM to give the MSSM, there is an extensive literature on
``gauged'' $R$-parity, \ie~where $R$-parity is the remnant of a broken
gauge symmetry. Martin has considered $R$-parity as embedded in a
$U(1)_{{B-L}}$ gauge symmetry and classified the possible order
parameters in extended gauge symmetries [$SO(10)$, $SU(5)$,
$SU(5)\times U(1)$, $E_6$], which necessarily lead to $R$-parity
\cite{Martin:1992mq,Martin:1996kn}. Babu \textit{et al.}
\cite{Babu:2002tx} combine DGSs with an attempt to solve the
$\mu$-problem. Chemtob {\it et al.} \cite{Chemtob:2006ur} deal with 
anomaly-free DGSs of the next-to-MSSM (NMSSM).
Although not in our systematic context, some of the
anomaly-free DGSs we find are mentioned in the literature explicitly
\cite{Babu:2002tx} or implicitly \cite{Hamaguchi:1998wm}. In
particular, $\bsym{P}_6$ occurs in Ref.~\cite{Babu:2002tx}, and in
Refs.~\cite{Babu:2003qh,Wang:2004mg} a related non-supersymmetric
$\maZ_6$ is studied.


\section{The Linear Anomaly Constraints}
\label{linear}
\cleqn
In this section, we review the work of IR
\cite{Ibanez:1991hv,Ibanez:1991pr} on DGSs. We focus here on
constraints arising from the linear $U(1)_X$ anomalies $\mc{A}_{CCX}$,
$\mc{A}_{WWX}$ and $\mc{A}_{GGX}$, where we adopt the notation of
Ref.~\cite{Dreiner:2003yr}. For example, the $SU(3)_C$-$SU( 3)_C$-$U(1)_X$ 
anomaly is denoted as $\mc{A}_{CCX}$, and $G$ stands for
``Gravity''. In Sect.~\ref{cubic}, we investigate the purely Abelian
anomalies, {\it i.e.}~$\mc{A}_{YYX}$, $\mc{A}_{YXX}$ and especially
the cubic anomaly $\mc{A}_{XXX}$.

For the high-energy gauge symmetry, we consider an in general
generation-dependent $U(1)_X$ extension of $\Gsm$, with the chiral
superfield charges quantised (\emph{i.e.} the quotient of any two
charges is rational) and normalised to be integers. We assume it is
spontaneously broken by the vacuum expectation value (VEV), $\ups$, of
a scalar field $\Phi$ with $U(1)_X$~charge $X_\Phi \equiv N>1$. The
mass scale of the broken symmetry is $M_X={\cal O}(\ups)\gg M_W$. (We
assume here a single field $\Phi$, or a vector-like pair; {\it cf.} 
Sect.~\ref{gauged}.)  This leaves a residual, low-energy $\maZ_{N} $
symmetry, which we assume to be generation-independent\footnote{Note that due 
to the {\it three} non-vanishing mixing angles of the CKM matrix, 
one is forced to work with generation-independent discrete charges 
for the quarks. Concerning the leptons, generation-dependence is 
only possible if one relies on radiatively generated neutrino masses. 
See Ref.~\cite{Kapetanakis:1992jj}.} on the SSM
chiral superfields \cite{Krauss:1988zc,Banks:1989ag}. In the
low-energy theory, we restrict ourselves to the particle content of
the SSM, allowing however for additional heavy fermions with masses
${\cal O}(M_X)$. To avoid later confusion, we emphasise here that the
$U(1)_X$ charge of $\Phi$ is not necessarily the same $N$, which
appears in the final $\maZ_N$ we obtain when restricting ourselves to
the so-called ``fundamental'' DGSs. We discuss this in more detail in
Sect.~\ref{rescaling}.

For the SSM fields, the $\maZ_N$ charges $q_i$ are related to the
integer $U(1)_X$ charges, $X_i$, via a modulo $N$ shift
\beq
X_i=q_i + m_i N\,.\label{charges}
\eeq
Here the index $i$ labels the SSM particle species and $q_i,\,m_i$ are
integers.  Just like the $U(1)_X$ charges, the $m_i$ are in general
generation-\emph{dependent}, whereas the $q_{i}$ are assumed to be
generation-independent. We also allow for Dirac and Majorana fermions
which become massive at ${\cal O}(M_X)$. For the former, two fields
with $U(1)_X$ charges $X^j_{\mathrm{D1}}$ and $X^j_{\mathrm {D2}}$,
respectively, must pair-up, resulting in a Dirac mass term after
$U(1)_X$ breaking. The Majorana fields with charge $X^{j^\prime}_
{\mathrm {M}}$ can directly form a mass term. The $\maZ_N$ invariance
of these mass terms requires
\beqn
X^j_{{\mathrm{D1}}} + X^j_{{\mathrm{D2}}} &=& 
p_j N, \qquad p_j \in\mathbb{Z},\label{dirac} \\ [0.3cm]
2\cdot X^{j^\prime}_{\mathrm{M}} &=& 
p_{j^\prime}^{\prime} N,\qquad p_{j^\prime}^{\prime} \in \mathbb{Z}\,.
\label{majorana} 
\eeqn
The indices $j$ and $j^\prime$ run over all heavy Dirac and Majorana
particles, respectively.

Assuming the initial $U(1)_X$ is anomaly-free, IR derive the resulting
constraints on the $\maZ_N$ charges $q_i$ of Eq.~(\ref{charges}). From
the anomaly cancellation conditions $\mc{A}_{CCX}=\mc{A}_{WWX}=\mc{A}
_{GGX}=0$, we obtain
\beqn
\sum_{i=\mathbf{3},\overline{\mathbf{3}}} q_i & = & - N\!\cdot\! 
\left[ \sum_{i=\mathbf{3},\overline{\mathbf{3}}} m_i + \sum_
{j=\mathbf{3},\overline{\mathbf{3}}} p_j\right],\label{ccx}\\
\sum_{i=\mathbf{2}} q_i & = & - N\!\cdot \!\left[\sum_{i=\mathbf{2}} m_i 
+ \sum_{j=\mathbf{2}} p_j\right], \label{wwx}\\
\sum_i q_i &=& - N\! \cdot\!\left[ \sum_i m_i + \sum_j p_j + \sum_{j^\prime} 
\mbox{$\frac{1}{2}$}p_{j^\prime}^{\prime}\right]\,.\label{ggx}
\eeqn
The sums in Eqs.~(\ref{ccx}) and (\ref{wwx}) run over all colour
triplets and weak doublets, respectively, \textit{i.e.}~we restrict
ourselves to only fundamental representations\footnote{The
contribution of a fermion to an $SU(M)$-$SU(M)$-$U(1)$ anomaly is
proportional to the corresponding Dynkin index
\cite{Slansky:1981yr}. Particles constituting higher irreducible
representations of $SU(M)$ have a Dynkin index which is an integer
multiple of that of the fundamental $M$-plet
\cite{Babu:2002tx,Wang:2004mg}. Therefore heavy particles in higher
irreducible representations need not be considered for our purposes,
see Eqs.~(\ref{ccxZ})-(\ref{ggxZ}). Note that in Eqs.~(\ref{ccx}) and
(\ref{wwx}) we do not consider Majorana particles either, because all
real representations of $SU(M)$ have a Dynkin index which is an even
multiple of that of the fundamental irreducible representation, see
Refs. \cite{Babu:2002tx,Banks:1991xj}.} of $SU(3)_C$ and $SU(2)_W$. As all
particles couple gravitationally, we sum over the entire chiral
superfield spectrum in Eq.~(\ref{ggx}).

Depending on the charge shifts, $m_i$, of the low-energy fields, as well as
the heavy-fermion particle content, the square brackets in Eqs.
(\ref{ccx})-(\ref{ggx}) can take on arbitrary integer values. In the
case of even $N$, any half-odd integer is allowed for the square bracket in
Eq. (\ref{ggx}). Hence, we can rewrite them symbolically as
\beqn
\sum_{i=\mathbf{3},\mathbf{\overline{3}}} q_i & = &  
N \cdot \mathbb{Z},\label{ccxZ}\\
\sum_{i=\mathbf{2}} q_i & = &  N \cdot \mathbb{Z}, \label{wwxZ} \\
\sum_i q_i &=&  N \cdot  \mathbb{Z} + \eta \cdot\frac{N}{2}
\cdot \mathbb{Z},\label{ggxZ}
\eeqn
with $\eta=0,1$ for $N=\mathrm{odd,even}$, respectively. From the
point of view of the low-energy theory, the various $\mathbb{Z}$s,
including the two in Eq.~(\ref{ggxZ}), \textit{each} represent an
arbitrary and independent integer, which is fixed by the heavy-fermion
content and the choice of $m_i$.

In addition to the anomaly constraints, we obtain constraints on the
$U(1)_X$ charges, by requiring a minimal set of interaction terms in
the SSM superpotential, which are responsible for the low-energy
fermion masses, namely the first three terms in Eq.~(\ref{superpot}).
In Sect.~\ref{physics} we investigate the consequences of additionally
imposing $H_dH_u$ invariance.  The $\maZ_N$ charge equations
corresponding to the first three terms of Eq.~(\ref{superpot}) are
\beqn
q_L+q_{H_d}+q_{\bar{E}}&=&0~\mbox{mod}~N\,,\label{f-masses1} \\
q_Q+q_{H_d}+q_{\bar{D}}&=&0~ \mbox{mod}~N\,, \label{f-masses2}\\
q_Q+ q_{H_u}+q_{\bar{U}}&=&0~\mbox{mod}~N\,. \label{f-masses3}
\eeqn
These are three equations for seven unknowns. We can thus write the
family-independent $\maZ_N$ charges of the SSM superfields 
in terms of four independent integers, which we choose as $m,n,p,r=0
,1,...,N-1$.
\beqn
q_Q\!\!\!&=&\!\!\! r, \qquad \qquad \qquad \qquad
q_{\bar{U}}=-m-4r, \qquad \;\;q_{\bar{D}}=m-n+2r\,,
\notag \\
\quad q_L\!\!\!&=&\!\!\!-n-p-3r, \qquad \quad\, q_{\bar{E}}=m+p+6r, \qquad 
 \notag \\
q_{H_d}\!\!\!&=&\!\!\!
-m+n-3r, \qquad \,\, q_{H_u}=m+3r\,.
\label{r-eqs}
\eeqn
In the following, we make use of the integer normalised hypercharges
\beq
Y(Q,{\bar U},{\bar D},L,{\bar E},H_d,H_u) = (-1,4,-2,3,-6,3,-3)\,.
\label{hyper}
\eeq
The choice of integers $m,n,p$ in Eq.~(\ref{r-eqs}) corresponds to the
notation of IR. The slightly unusual coefficients for the integer $r$
correspond to the negative normalised hypercharge given in
Eq.~(\ref{hyper}), and were chosen for the following charge
transformation: To simplify the up-coming calculations, we perform a
shift of the integer $\maZ_N$ charges by their integer hypercharges,
such that the resulting charge ${q_Q}'$ is zero,
\beq
q_i\longrightarrow {q_i}^\prime=q_i+Y_i\cdot r\,.
\label{y-shift}
\eeq
In the following, we drop the prime on the charge symbols. This shift
in the $\maZ_N$ charges does \textit{not} change the effect of
$\maZ_N$ on the renormalisable or non-renormalisable operators of the
SSM superpotential or $D$-terms, since these are all $U(1)_Y$
invariant. It also does not affect the anomaly-equations which we
consider. However, it does correspond to a change in the underlying
$U(1)_X$ gauge theory. The difference can lead to in principle
observable effects, for example cross-sections which depend on
$X$-charges. We return to this change in Sect.~\ref{top-down}.

The choice of charges where $q_Q=0$, is the basis in which IR
work. They show that in this case, any $\maZ_N$~symmetry $g_N$ can be
expressed in terms of the product of powers of the three (mutually
commuting) generators $R_N$, $A_N$ and $L_N$ \cite{Ibanez:1991pr}:
\beq
g_N\; =\; R_N^m \times A_N^n \times L_N^p,\quad
\mbox{with the exponents}\quad m,n,p=0,1,...N-1\,.
\label{elem-gen}
\eeq
The charges of the SSM chiral superfields under the three independent
$\maZ_N$~generators are given in Table~1 of Ref.~\cite{Ibanez:1991pr}.
In terms of the powers $m,n,p$, the generation-independent $\maZ_N$
charges of the SSM superfields are\footnote{The action of $g_N$ on
  {\it e.g.} the chiral superfields $\bar{D}_i$ is thus given by
  $\bar{D}_i \rightarrow \exp{\left[\frac{2\pi \mathrm{i}}{N}
      (m-n)\right]\:\bar{D}_i}.$}
\beqn
q_Q\!\!\!&=&\!\!\! 0, \qquad \qquad \qquad\;
q_{\bar{U}}=-m, \qquad \quad q_{\bar{D}}=m-n
\notag \\
\quad q_L\!\!\!&=&\!\!\!-n-p, \qquad \quad\;\;\, q_{\bar{E}}=m+p, \qquad 
 \notag \\
q_{H_d}\!\!\!&=&\!\!\!
-m+n, \qquad \quad q_{H_u}=m\,.
\label{discharges}
\eeqn
Note that the integers $m,n,p$ here are the same as in
Eq.~(\ref{r-eqs}). Inserting the charges above into
Eqs.~(\ref{ccxZ})-(\ref{ggxZ}), and assuming the SSM light-fermion
content we arrive at the conditions\footnote{These equations are
$r$-independent, they result by directly plugging Eq.~(\ref{r-eqs})
into Eqs.~(\ref{ccxZ})-(\ref{ggxZ}). However, when considering the
cubic anomaly in Sect.~\ref{cubic}, the $r$-dependence does not
cancel.}
\beqn
3 n &=&   N \cdot \mathbb{Z}\,,\label{ccxZZ}\\
3(n+p)-n  &=&   N \cdot \mathbb{Z}\,, \label{wwxZZ} \\
3(5\,n+p-m)-2n&=&  N \cdot  \mathbb{Z} + \eta \cdot\mbox{$\frac{N}{2}$}
\cdot \mathbb{Z}\,.\label{ggxZZ}
\eeqn
Since all $\mathbb{Z}$s in Eqs.(\ref{ccxZZ})-(\ref{ggxZZ}) stand for
arbitrary and independent integers, we can combine these Diophantine
equations to obtain a simpler set,
\beqn
3 n & = &  N \cdot \mathbb{Z}\,,\label{l1}\\
3p-n & = &  N \cdot \mathbb{Z}\,, \label{l2} \\
3(m+p)&=&  N \cdot  \mathbb{Z} + \eta \cdot\mbox{$\frac{N}{2}$}\cdot 
\mathbb{Z}\,.\label{l3}
\eeqn
This differs slightly from IR in notation, as we find it more
convenient to retain the arbitrary integers $\mathbb{Z}$ on the
right-hand side.  These three equations are the basis for our further
study. DGSs satisfying all three equations will be called
``anomaly-free DGSs'', although these constraints are only necessary
but not sufficient for complete anomaly-freedom of the high-energy
theory \cite{Banks:1991xj,Ibanez:1992ji}.


\section{Symmetries Allowed by the Linear Constraints}
\label{linear-2}
\cleqn
In this section, we go beyond the work of IR and determine the
solutions, $(n,p,m;N)$, to the Eqs.~(\ref{l1})-(\ref{l3}) for
\textit{general} values of $N$, not just $N=2,3$. We separately
consider the two possibilities: either $N$ \textit{is not} or
\textit{is} a multiple of $3$. We employ the notation:
\beqn
(k|N)&:\Leftrightarrow& N=0\, {\rm mod}\;k\,,\notag \\ 
\neg(k|N) &:\Leftrightarrow& N\not=0\,{\rm mod}\;k \,. \notag
\eeqn
$k\in\mathbb{N}$, where $\mathbb{N}$ is the set of all positive 
integers including zero.

\begin{enumerate}
\item {\bf $\bsym{\neg \,(3 \,| N)}$:} Since $n=0,1,...,N-1$,
  Eq.~(\ref{l1}) requires $n=0$. Then Eq.~(\ref{l2}) similarly gives
  $p=0$. Finally, Eq.~(\ref{l3}) then implies 
\begin{enumerate}
\item $m=0$ for odd $N$. This is the case of the trivial symmetry, the
  identity.
\item For even $N$ there are two possibilities, either $m=0$ (trivial) or 
$m=\frac{N}{2}$.
\end{enumerate} 
We conclude that the only {\it non-trivial} anomaly-free DGSs here are
\beqn
g_N = R_N^{N/2},\quad N=\mathrm{even}\,.\label{rp}
\eeqn
The simplest case with $N=2$ yields the discrete $\maZ_2$ charges:
$q_Q=q_L=0$, $q_{\bar {D}}=q_{\bar{E}}=q_{H_u}=1$, $q_{\bar{U}}=q_
{H_d}=-1$. This charge assignment is, from the low-energy point of
view, equivalent to standard matter-parity
\cite{Bento:1987mu}. A reversed hypercharge shift,
Eq.~(\ref{y-shift}), back to Eq.~(\ref{r-eqs}) with $r=1$ yields:
$q_Q=q_L=q_{\bar {D}}=q_{\bar {U}}= q_{\bar{E}}=1\,{\rm mod}\, 2$,
$q_{H_u}=q_{H_d}=0$.

\item {\bf $\bsym{(3\, | N)}$:} Here we can define an $N'\in
  \mathbb{Z}$, such that $N\equiv 3 N'$.  From Eq.~(\ref{l1}) we
  obtain $n=0,N',$ or $2N'$:
\begin{enumerate}
\item Focusing first on $n=0$, we see that $p=\ell_p N'$, for
  $\ell_p=0,1,2$. Concerning Eq.~(\ref{l3}), it is again necessary to
  distinguish between odd and even $N$. Thus we find a set of
  anomaly-free DGSs
\beq
n=0,~~ p=\ell_p N',~~ m=\left\{ \begin{array}{ll} \ell_m N', & 
~ N=\mathrm{odd}, \\ 
s_m \frac{N'}{2}, & ~ N=\mathrm{even}, 
\end{array} \right.\label{n=0}
\eeq
with $\ell_p,\,\ell_m=0,1,2$ and $s_m=0,1,...,5$.

\item Inserting $n=N'$ into Eq.~(\ref{l2}), we obtain $p=\frac{N'}{3}
  +\ell_p N'$, again with $\ell_p=0,1,2$. For $p\in\mathbb{Z}$, we
  need $(3\,|N')$ or equivalently $N'\equiv 3N''$, with $N''\in
  \mathbb{Z}$. Taking into account Eq.~(\ref{l3}), we now find
\beq
n=N',  p= (1 + 3\,\ell_p) N''\,, \;\;m=\left\{ 
\begin{array}{ll} 
(2+ 3\, \ell_m)\, N'', & ~ N=\mathrm{odd},\\ 
(1 + 3\,s_m)\, \frac{N''}{2}, & ~ N=\mathrm{even}\,. 
\end{array} \right.
\label{n=N'}
\eeq

\item Analogously, $n=2N'$ gives
\beq
n=2N',~~p=(2 + 3\,\ell_p) N'',~~m=\left\{ \begin{array}{ll}  
(1 + 3\,\ell_m) N'', & ~ N=\mathrm{odd}, \\ (2 + 3\,s_m )
\frac{N''}{2}, & ~ N=\mathrm{even}.\end{array}   \right.\label{n=2N'}
\eeq

\end{enumerate}

The class of DGSs given in (c) need not be investigated any further
for it is equivalent to the one in (b): A $\maZ_N$~symmetry with
charges~$q_i$ is indistinguishable from one with charges~$-q_i$;
therefore the sets $(n,p,m)$ and $(N-n,N-p,N-m)$ yield equivalent
DGSs. {As an example, consider the integer $p$. For every $p_2$ in
Eq.~(\ref{n=2N'}) require a $p_1$ in Eq.~(\ref{n=N'}), such that
$p_2\stackrel {!}{=} N-p_1$. Inserting Eqs.~(\ref{n=2N'}) and
(\ref{n=N'}), we obtain $(2 + 3\ell_ {p_2}) N''\stackrel{!}{=}(9 - 1 -
3 \ell_{p_1}) N''$, which is solved for $\ell_{p_1}=2-\ell_{p_2} \in\{
0,1,2\}$. Similarly, the integer $m$ can be treated for even or odd
$N$.}  Likewise, some DGSs of Eq.~(\ref{n=0}) are not independent of
the others.
\end{enumerate}

\begin{table}[t]

\hspace{0.7cm} \begin{tabular}{||c|l||c|c|c|c||} 
\hline \hline 
\multicolumn{2}{||c||}{\rule[-3mm]{0mm}{8mm}$\maZ_N$ Category} & $n$ & 
$p$ & $m$ & \# indep. $g_N$ \\ 
\hline \hline 
\rule[-3mm]{0mm}{8mm} $\neg (3\,|N)$ & $N$ even  & $0$ & $0$ & $\frac{N}{2}$ 
& 1 \\ 
\hline \hline 
\rule[-3mm]{0mm}{8mm}  & $N$ odd  & $0$ & $(0,1) \cdot N'$ & $(0,1,2)\cdot N'$
 & 4 \\
\cline{2-6}
 \rule[-3mm]{0mm}{8mm}  \raisebox{2.2ex}[-2.2ex]{$\phantom{\neg}(3\,|N)$} & 
$N$ even  & $0$ & $(0,1) \cdot N'$ & $(0,1,2,3,4,5)\cdot \frac{N'}{2}$ & 9 \\
\hline 
\rule[-3mm]{0mm}{8mm} & $N$ odd  & $N'$ & $(1,4,7) \cdot N''$ & $(2,5,8)
\cdot N''$ & 9 \\
\cline{2-6}
\rule[-3mm]{0mm}{8mm}   \raisebox{2.2ex}[-2.2ex]{$\phantom{\neg}(9\,|N)$} & 
$N$ even  & $N'$ & $(1,4,7) \cdot N''$ & $(1,4,7,10,13,16)\cdot 
\frac{N''}{2}$ 
& 18 \\
\hline \hline
\end{tabular} \caption{\small The list of all DGSs
satisfying the linear anomaly constraints of Ib\'a\~nez and Ross. $N'$
and $N''$ are defined by $N=3N'=9N''$, where $N,N',N''\in\mathbb{N}$.
The $\ell_p=2$ cases are not listed as they are equivalent to the set
of DGSs with $\ell_p=1$. The last column gives the resulting number of
independent non-trivial DGSs, $g_N$, for fixed $N$.}\label{lintab}
\end{table}

Table~\ref{lintab} summarises the anomaly-free DGSs classified by $N$
and the powers $n$, $p$ and $m$. For example, the two rows with
$(3\,|N)$ correspond to the DGSs of Eq.~(\ref{n=0}). The last column
shows the number of independent non-trivial $g_N$.  The $4$ in the
second row arises because there are three DGSs with $\ell_p=1$ but
only one with $\ell_p=0$; with $p=0$, the case $m=0$ is trivial,
whereas $m=N'$ and $m=2N'$ lead to equivalent DGSs. Similarly, we get
$9$ DGSs instead of $12$ for the third row.


\section{The Purely Abelian Anomalies}
\label{cubic}
\cleqn
So far, we have determined the constraints on DGSs arising from the
three linear anomaly conditions of Eqs.~(\ref{ccx})-(\ref{ggx}). Next
we consider the three purely Abelian anomalies $\mc{A}_{YYX}$,
$\mc{A}_{YXX}$ and $\mc{A}_{XXX}$, respectively.

\begin{enumerate}

\item Analogously to Eqs.~(\ref{ccx})-(\ref{ggx}), we obtain from $\mc
{A}_{YYX}=0$ that
\beq\label{yyx}
\sum_{i} {Y_i}^2~q_i~=~-N\left[
\sum_i {Y_i}^2~m_i~+~\sum_j {Y_{\mathrm{D1}}^j}^2~p_j\right].
\eeq 
We have used $Y_{\mathrm{D2}}^j=-Y_{\mathrm{D1}}^j$ and
$Y_{\mathrm{M}}^j=0$, as well as Eq.~(\ref{dirac}). Note that each
term, unlike those in Eqs.~(\ref{ccx})-(\ref{ggx}), contains a factor
of ${Y_{...}}^2$, which is in general different for each
field.\footnote{In the case of the non-Abelian linear anomalies
$\mathcal{A}_{CCX}$ and $\mathcal{A} _{WWX}$, one encounters a factor
proportional to the Dynkin index instead. This is a common factor for
all fields provided they are all in the fundamental representation of
$SU(3)_C$ and $SU(2)_W$, respectively.}  Recall, that we have chosen
the hypercharges to be integer for all SSM particles, see
Eq.~(\ref{hyper}). Thus the left-hand side is integer. However, given
this normalisation, the hypercharges of the heavy fermions need not be
integer and the quantity in square brackets need not be in
$\mathbb{Z}$. Thus the right-hand can take on {\it any} value within
$\mathbb{Z}$. Therefore Eq.~(\ref{yyx}) poses no constraint.

\item Now $\mathcal{A}_{YXX}=0$. Analogously to Eq.~(\ref{yyx}), we
  get \beq\label{yxx} \sum_{i} {Y_i}~{q_i}^2~=~-N\left[ \sum_i
  {Y_i}~m_i({m_i}~N+2q_i)~-~\sum_j {Y_{\mathrm{D1}}^j}~p_j (p_jN-2
  X_{\mathrm{D1}}^j)\right].  \eeq By considering only the
  $Y_{\mathrm{D1}}^j$, we see that $[...]$ is not necessarily an
  integer, just as in the previous case. Thus Eq.~(\ref{yxx}) is of no
  use from the low-energy point of view.\footnote{We disagree here
  with Refs.~\cite{Ibanez:1991hv,Ibanez:1991pr} about the reason why
  $\mathcal{A}_{YYX}$ and $\mathcal{A}_{YXX}$ do not impose useful
  constraints on $\maZ_N$~symmetries. It is {\it not} the (overall)
  normalisation of the Abelian charges, but the fact that these
  charges are in general different for each field.}

\item Next, we consider the cubic anomaly $\mathcal{A}_{XXX}$. Here we 
do not have a mixture of known and unknown charges: We do not know any
of the $U(1)_X$ charges. We obtain for the anomaly-equation
\beqn
\sum_i {q_i}^3  &=& - \sum_i \left(3{q_i}^2m_iN + 3q_i{m_i}^2N^2 +{m_i}^3N^3
   \right) \notag \\ 
&& - \sum_j \left( 3 {X^j_{\mathrm{D1}}}^2 p_j N - 3 {X^j_{\mathrm
{D1}}}{p_j}^2 N^2 +{p_j}^3 N^3 \right) \notag \\ 
&& - \mbox{$\frac{1}{8}$}\sum_{j^\prime}  
{p_{j^\prime}^{\prime}}^3 N^3\,.\label{cub}
\eeqn
{\it If} fractional $X^j_{\mathrm{D1}}$ were allowed, again no
extraction of a meaningful constraint is feasible, since in this case
the right-hand side of Eq.~(\ref{cub}) is not necessarily of the
form $N\cdot\mathbb{Z}$. However, as outlined in Sect.~\ref{linear},
we only consider integer $X$-charges here. We shall investigate the
case of fractional $X$-charges for the heavy fields in
Sect.~\ref{rescaling}, since the difference can be meaningful in
cosmology
\cite{Athanasiu:1988uj,Ellis:1990iu,Ellis:2004cj}. 

The calculation for the cubic anomaly with only integer charges is
similar to the calculation in Sect.~\ref{linear-2}, \ie\ it involves
many case distinctions. It can be found in Appendix~\ref{app-cubic}.
In Table~\ref{cubtab}, we have summarised the results. We show those
$N$, as well as the powers $(n,p,m)$, in the case of only integer
$X$-charges, which satisfy both the linear anomaly constraints of
Sect.~\ref{linear-2} (\cf~Table~\ref{lintab}), as well as the cubic
anomaly equation considered here.
\begin{table}[t]

\hspace{0.6cm} \begin{tabular}{||c|l||c|c|c|c||} 

\hline \hline 

\multicolumn{2}{||c||}{\rule[-3mm]{0mm}{8mm}$\maZ_N$ Category} & $n$ & 
$p$ & $m$ &  possible $N$ \\ 

\hline \hline 

\rule[-3mm]{0mm}{8mm} $\neg (3\,|N)$ & $N$ even  & $0$ & $0$ & $\frac{N}{2}$ 
& $2\cdot\mathbb{N}$ \\ 

\hline \hline

\rule[-3mm]{0mm}{8mm}  &   & $0$ & $(0,1) \cdot N'$ & $(0,1,2)\cdot N'$ & 
$9\cdot(2\!\cdot\!\mathbb{N}+1)$ \\

\cline{3-6}

\rule[-3mm]{0mm}{8mm}  & \raisebox{2.2ex}[-2.2ex]{$N$ odd}  & $0$ & $  N'$ & 
$ N'$ & $3\cdot(2\!\cdot\!\mathbb{N}+1)$ \\

\cline{2-6}

\rule[-3mm]{0mm}{8mm}   &   & $0$ & $(0,1) \cdot N'$ & $ (0,1,2,3,4,5)\cdot 
\frac{N'}{2}$ &  $18\cdot\mathbb{N}$  \\

\cline{3-6}

\rule[-3mm]{0mm}{8mm}  &   & $0$ & $0$ & $  \frac{N}{2}$ &  
$6\cdot\mathbb{N}$  \\

\cline{3-6}

\rule[-3mm]{0mm}{8mm}  &   & $0$ & $ N'$ & $ N'$ &  $6\cdot\mathbb{N}$  \\

\cline{3-6}

\rule[-3mm]{0mm}{8mm}  \raisebox{15.3ex}[-15.3ex]{$\phantom{\neg}(3\,|N)$} & 
\raisebox{6.8ex}[-6.8ex]{$N$ even}  & $0$ & $ N'$ & $ 5\cdot \frac{N'}{2}$ &  
$6\cdot\mathbb{N}$  \\

\hline \hline

\rule[-3mm]{0mm}{8mm} & $N$ odd  & $N'$ & $(1,4,7) \cdot N''$ & $(2,5,8)\cdot 
N''$ &  $27\cdot(2\!\cdot\!\mathbb{N}+1)$ \\

\cline{2-6}

\rule[-3mm]{0mm}{8mm}   \raisebox{2.2ex}[-2.2ex]{$\phantom{\neg}(9\,|N)$} & 
$N$ even  & $N'$ & $(1,4,7) \cdot N''$ & $(1,4,7,10,13,16)\cdot \frac{N''}{2}$
 & $54\cdot\mathbb{N}$ \\

\hline \hline

\end{tabular} \caption{\small Compatibility of the linear and the cubic 
anomaly constraints in the case of integer $U(1)_X$ charges for {\it
all} chiral superfields. For each $\maZ_N$ category, the allowed
values of $N$ are given in the far right column. The DGSs are
specified by the set $(n,p,m)$, in accordance with 
Eq.~(\ref{elem-gen}). We employ the notation: $N'\equiv N/3$, $N''
\equiv N/9$, and $N',\,N''\in\mathbb{N}$. For special values of $N$, 
all linearly allowed DGSs are compatible with the cubic anomaly
condition. However, four classes of DGSs within the categories 
$(3\,|N)$ (rows 3, 5, 6, 7) are possible for less constrained
$N$.}
\label{cubtab}
\end{table}
The main effect of the cubic anomaly constraint consists in reducing
the (infinite) list of possible DGSs. Considering $N=3$ for instance,
there are four independent $g_N$ symmetries allowed in
Table~\ref{lintab}. However, only one of these, namely the case where
$(n,p,m)=(0,1,1)$, complies with Table~\ref{cubtab}. This corresponds
to $\bsym{B}_3$, \ie~baryon-triality discussed by IR.

Another example is $N=6$. Here we have nine linearly allowed DGSs,
while only three are left after imposing the cubic anomaly constraint:
$R_6^3$, $R_6^2L_6^2$ and $R_6^5L_6^2$. The first two are physically
equivalent to $\bsym{M}_p$ and $\bsym{B}_3$ from the low-energy point
of view. We shall denote $\bsym{P}_6\equiv R_6^5L_6^2$, as
proton-hexality. This is a special discrete symmetry, which we return
to in Sect.~\ref{physics}. For $N=9$ there are $4+9$ linearly allowed
$g_N$, of which only four are also consistent with the cubic anomaly
condition.  $N=27$ is the first case for $(3|N)$, where the cubic
anomaly does not reduce the number of allowed DGSs.

\end{enumerate}


\section{Charge Rescaling}
\label{rescaling}
\cleqn

So far, we have assumed that hypercharge shifted discrete symmetries,
as in Eq.~(\ref{y-shift}), are equivalent and {\it all} chiral
superfields have integer $U(1)_X$ charges.  However, from the
\textit{low-energy} point of view, this latter assumption is too
restrictive \cite{Banks:1991xj,Ibanez:1992ji}. To see this in our
analysis, consider an example from Table~\ref{cubtab}, where
$N=18$. The powers of the elementary discrete gauge group generators,
Eq.~(\ref{elem-gen}), are given by
\beq
n=0,\qquad p=6\cdot (0,1),\qquad
m=3\cdot s_m,\quad s_m=0,1,\ldots,5\,,
\eeq
which are all multiples of the common factor $F=3$. The charges of the
SSM fields, $q_i + m_i N$, are given in Eq.~(\ref{discharges}) as
linear combinations of $n,\,p,$ and $m$, and are therefore also all
multiples of $F$, in our example. From the low-energy point of view,
with the heavy fields integrated out, such a charge assignment is
indistinguishable from a scaled one with charges $(q_i+m_iN)/F$. After
the breakdown of $U(1)_X$, the residual DGS is then a $\maZ _{N/F}$
instead of a $\maZ_{N}$. However, the $\maZ_{N/F}$ does not
necessarily satisfy the cubic anomaly, with all integer charges. In
our example, we have $N/F=6$, which, according to Table~\ref{cubtab},
satisfies the cubic anomaly only for very special values of $(n,p,m)$.

This integer rescaling only applies to the charges of the SSM chiral
superfields. For the \textit{heavy} fermions, it is typically not
possible and leads to fractional charges. From a bottom-up approach,
experiments would determine the rescaled DGS group $\maZ_{N /F}$. When
searching for the possible (low-energy) anomaly-free DGSs, we
therefore relax our original assumption of integer charges and instead
allow fractional charges for the heavy sector, only. We then denote
the DGS $\maZ_{N/F} $ with the {\it maximally} rescaled charges as the
\textit{fundamental} DGS, {\it i.e.} $F$ is the largest common factor
of $N$ and all $q_i + m_i N$. In Table~\ref{fundtab}, we present the
complete list of fundamental DGSs, obtained from
Table~\ref{cubtab}. We see that after rescaling, the infinite number
of DGSs listed in Table~\ref{cubtab} is reduced to a finite set of 27
fundamental $\maZ_N$~symmetries: one with $N =2$, four with $N=3$,
four with $N=6$, nine with $N=9$ and nine with $N=18 $.

\begin{table}[t]

\hspace{2.3cm} \begin{tabular}{||c||c|c|c|l||} \hline \hline 

{\rule[-3mm]{0mm}{8mm}$N$} & $n$ & $p$ & $m$ & DGSs \\\hline\hline 

\rule[-3mm]{0mm}{8mm} $2$ & $0$ & $0$ & $1$& 
$\!$ \underline{$R_2^{\phantom{1}}$}\\\hline\hline
\rule[-3mm]{0mm}{8mm}   & $0$ & $ 0 $ & $1$ & $R_3^{\phantom{1}}$\\ 
\cline{2-5}

\rule[-3mm]{0mm}{8mm}   \raisebox{2.1ex}[-2.1ex]{$3$}  & $0$ & $ 1 $ & 
$(0,1,2)$ & $L_3^{\phantom{1}},
%
%
%
%
\:$\underline{$L_3^{\phantom{1}}R_3^{\phantom{1}}$}$,~L_3^
{\phantom{1}}R_3^2$\\ \hline \hline

\rule[-3mm]{0mm}{8mm}   & $0$ & $ 0 $ & $1$ & $R_6^{\phantom{1}}$\\
\cline{2-5}

\rule[-3mm]{0mm}{8mm}   \raisebox{2.1ex}[-2.1ex]{$6$}  & $0$ & $ 2 $ & 
$(1,3,5)$ & $L_6^2R_6^{\phantom{1}},~L_6^2R_6^3,\:$\underline{$L_6^2R_6^5$} 
\\ \hline \hline
\rule[-3mm]{0mm}{8mm}   & $3$ & $ 1 $ & $(2,5,8)$ & $A_9^3L_9^{\phantom{1}}
R_9^2,~A_9^3L_9^{\phantom{1}}R_9^5,~A_9^3L_9^{\phantom{1}}R_9^8$ \\
\cline{2-5} 

\rule[-3mm]{0mm}{8mm} $9$  & $3$ & $ 4 $ & $(2,5,8)$ & $A_9^3L_9^4R_9^2,~A_9^3
L_9^4R_9^5,~A_9^3L_9^4R_9^8$ \\ \cline{2-5}

\rule[-3mm]{0mm}{8mm}   & $3$ & $ 7 $ & $(2,5,8)$ & $A_9^3L_9^7R_9^2,~A_9^3
L_9^7R_9^5,~A_9^3L_9^7R_9^8$ \\

\hline\hline

\rule[-3mm]{0mm}{8mm}  & $6$ & $ 2 $ & $(1,7,13)$ & $A_{18}^6L_{18}^2R_{18}^
{\phantom{1}},~A_{18}^6L_{18}^2R_{18}^7,~A_{18}^6L_{18}^2R_{18}^{13}$ \\
\cline{2-5}

\rule[-3mm]{0mm}{8mm} $18$  & $6$ & $ 8 $ & $(1,7,13)$ & $A_{18}^6L_{18}^8R_
{18}^{\phantom{1}},~A_{18}^6L_{18}^8R_{18}^7,~A_{18}^6L_{18}^8R_{18}^{13}$\\
\cline{2-5}

\rule[-3mm]{0mm}{8mm}  & $6$ & $ 14 $ & $(1,7,13)$ & $A_{18}^6L_{18}^{14}
R_{18}^{\phantom{1}},~A_{18}^6L_{18}^{14}R_{18}^7,~A_{18}^6L_{18}^{14}R_{18}^
{13}$ \\ \hline \hline
\end{tabular} 
\caption{\small All fundamental DGSs satisfying the linear and the 
cubic anomaly cancellation conditions. The heavy-fermion charges, 
$X^j$, are allowed to be fractional. The three underlined DGSs can 
be realised with only integer heavy-fermion $U(1)_X$ charges.}
\label{fundtab}
\end{table}

Refs.~\cite{Banks:1991xj,Ibanez:1992ji} pointed out that the cubic
anomaly-constraint is in general too restrictive on
\textit{low-energy} anomaly-free DGSs due to possible rescalings.
Comparing Table~\ref{cubtab} with Table~\ref{fundtab}, presents a
classification within the SSM of the solutions to this problem. As
emphasised earlier, the cubic anomaly constraint is compatible with
{\it all} five classes of linearly allowed DGSs presented in
Table~\ref{lintab}, however only for restricted values of
$N$. Rescaling the charges and allowing for fractionally charged heavy
fermions, eliminates the influence of the $\mathcal{A}_{XXX}$
condition on the fundamental DGSs completely. In other words,
\textit{all} linearly allowed \textit{fundamental} DGSs are compatible
with the cubic anomaly constraint. Therefore, Eq.~(\ref{cub}) contains
only information about whether or not the heavy-fermion
$U(1)_X$~charges are fractional or integer. Of the fundamental DGSs
listed in Table~\ref{fundtab}, solely $\bsym{M}_p
\equiv R_2^{\phantom{1}}$, $\bsym{B}_3\equiv R_3^{\phantom{1}}L_3^{
\phantom{1} }$ and $\bsym{P}_6\equiv R_6^5L_6^2$ are consistent with 
both the linear and the cubic anomaly conditions, without including
fractionally charged heavy particles.


\section{Physics of the Fundamental DGSs and the MSSM}
\label{physics}
\cleqn
Now that we have found a finite number of fundamental, anomaly-free
low-energy DGSs, we would like to investigate the correspondingly
allowed SSM operators. In particular, we study the effect of the 27
fundamental DGSs given in Table~\ref{fundtab} on the crucial baryon-
and/or lepton-number violating superpotential and K\"ahler potential
operators \cite{{Ibanez:1991pr},{Allanach:2003eb}}:
\beq
\begin{array}{lllclll}
{\cal O}_1&=&[LH_u]_F\,,& ~~~ & {\cal O}_2 & = & [LL\bar{E}]_F\,, \\ 
{\cal O}_3&=&[LQ\bar{D}]_F\,,& ~~~ & {\cal O}_4 & = & [\bar{U}
\bar{D}\bar{D}]_F\,, \\ 
{\cal O}_5&=&[QQQL]_F\,,& ~~~ & {\cal O}_6 & = & [\bar{U}
\bar{U}\bar{D}\bar{E}]_F\,, \\ 
{\cal O}_7 & = & [QQQH_d]_F\,, & ~~~ &   {\cal O}_8 & = & 
[Q\bar{U}\bar{E}H_d]_F\,, \\ 
{\cal O}_9 & = & [LH_uLH_u]_F\,, & ~~~ & {\cal O}_{10} 
& = & [LH_uH_dH_u]_F\,, \\ 
{\cal O}_{11} & = & [\bar{U}{\bar{D}}^\ast\bar{E}]_D\,, & ~~~ &  
{\cal O}_{12} & = & [{{H}_u}^\ast H_d\bar{E}]_D\,, \\ 
{\cal O}_{13} & = & [Q\bar{U}{L}^\ast]_D\,, & ~~~ &   {\cal O}_{14} & = & 
[QQ\bar{D}^\ast]_D\,.
\end{array} \label{dim5}
\eeq 
The subscripts $F$ and $D$ denote the $F$- and $D$-term of the
corresponding product of superfields. Table~\ref{phys} summarises
which operators are allowed for each fundamental anomaly-free DGS.
The symbol $\checkmark$ indicates that an operator is allowed. Thus,
for example, matter-parity ($R_2$) allows the operators $[H_dH_u]_F$,
but also the dimension-five baryon-number violating operators
$[QQQL]_F$ and $[\bar{U}\bar{U}\bar{D}\bar{E}]_F$, as well as the
lepton-number violating operators $[LH_uLH_u]_F$. We have included the
bilinear operators $LH_u$ (unlike IR), since, even under the most
general complex field rotation \cite{Dreiner:2003hw}, they can not be
eliminated, when taking into account the corresponding soft-breaking
terms \cite{Nardi:1996iy}.

\begin{table}[t]

\hspace{0.7cm} \begin{tabular}{||c||c||c|c|c|c||c|c|c|c||c||} 

\hline \hline 

{\rule[-3mm]{0mm}{8mm} } & $\!R_2^{\phantom{1}}\!$  &  $\!R_3^{\phantom{1}}L_
3^{\phantom{1}}\!$ & $\!R_3^{\phantom{1}}\!$  &  $\!L_3^{\phantom{1}}\!$ & 
$\!R_3^2L_3^{\phantom{1}}\!$    &$\!R_6^5L_6^2\!$&$\!R_6^{\phantom{1}}\!$&$
\!R_6^3L_6^2\!$ & $\!R_6^{\phantom{1}}L_6^2\!$  &  all $\maZ_9^{
\phantom{1}}\!$ \& $\!\maZ_{18}^{\phantom{1}}\!$\\ 

\hline \hline 

\rule[-3mm]{0mm}{8mm}$H_dH_u$  &$\checkmark$ &
$\checkmark$&$\checkmark$&$\checkmark$&$\checkmark$&$\checkmark$&$\checkmark$
&$\checkmark$&$\checkmark$&\\ \hline

\rule[-3mm]{0mm}{8mm}$L H_u$ &  &$\checkmark$&&&&&&& &\\\hline

\rule[-3mm]{0mm}{8mm}$LL\bar{E}$ &  &$\checkmark$&&&&&&& &\\\hline

\rule[-3mm]{0mm}{8mm}$LQ \bar{D}$ &  &$\checkmark$&&&&&&& &\\\hline

\rule[-3mm]{0mm}{8mm}$\bar{U} \bar{D}\bar{D}$ &  &&&$\checkmark$&&&&& 
&\\\hline

\rule[-3mm]{0mm}{8mm}$QQQL$ & $\checkmark$ &&$\checkmark$&&&&$\checkmark$&&&
\\\hline

\rule[-3mm]{0mm}{8mm}$\bar{U}\bar{U}\bar{D}\bar{E}$ &$\checkmark$  &&
$\checkmark$&&&&$\checkmark$&&  & \\\hline

\rule[-3mm]{0mm}{8mm}$QQQH_d $   &  &&&$\checkmark$&&&&& & \\\hline

\rule[-3mm]{0mm}{8mm}$Q\bar{U}\bar{E}H_d$   &  &$\checkmark$
&&&&&&& &\\\hline

\rule[-3mm]{0mm}{8mm}$LH_uLH_u$   &$\checkmark$ &
$\checkmark$&&&&$\checkmark$&&& &\\\hline

\rule[-3mm]{0mm}{8mm}$\!LH_uH_d H_u \!$  
& &$\checkmark$&&&&&&&& \\\hline \hline

\rule[-3mm]{0mm}{8mm}$\bar{U}{\bar{D}}^\ast\bar{E}$   & &$\checkmark$&&&&&&& &
 \\\hline

\rule[-3mm]{0mm}{8mm}${{H}_u}^\ast H_d\bar{E}$   & &
$\checkmark$&&&&&&& &\\\hline

\rule[-3mm]{0mm}{8mm}$Q\bar{U}{L}^*$  &  &$\checkmark$&&&&&&&  &\\\hline

\rule[-3mm]{0mm}{8mm}$QQ\bar{D}^*$     &  &&&$\checkmark$&&&&&  & \\

\hline \hline

\end{tabular} \caption{\small Physical consequences of the 27 
fundamental DGSs. The Higgs Yukawa couplings $LH_d\bar{E}$, $QH_ d\bar{D}$, 
and $QH_u\bar{U}$ are allowed for every DGS we
consider by construction. The symbol $\checkmark$ denotes that the
corresponding operator is possible for a given DGS. All anomaly-free
fundamental $\maZ_{9}$ and $\maZ_{18}$~symmetries forbid the operators
listed in the left column.}\label{phys}

\end{table}

We now demand the existence or absence of certain operators on
phenomenological grounds and thus further narrow down our choice of
DGSs.
\begin{itemize}
\item We have not included the term $[\mu\,H_dH_u]_F$ in the
  original list leading to Eqs.~(\ref{f-masses1})-(\ref{f-masses3}),
  since, in principle, it can be generated, {\it e.g.} dynamically
  \cite{Kim:1994eu,Giudice:1988yz,Chamseddine:1996rc,Harnik:2003rs}. From a
  low-energy point of view we must have $\mu\not=0$, and it must be of
  order the weak scale \cite{Drees:1996ca,lep}. There are attempts in
  the literature to combine the NMSSM or another dynamical mechanism to generate
  $\mu\not=0 $ with an anomaly-free DGS, see, for example,
  Ref.~\cite{Chemtob:2006ur}  or  Ref.~\cite{Babu:2002tx} 
 (and references therein), respectively. This is beyond the
  scope of this paper. If we explicitly require the $[\mu H_dH_u]_F
  $-operator in our theory, then as can be seen from Table~\ref{phys},
  all fundamental $\maZ_9^{\phantom {1}}$ and $\maZ_{18}^{
    \phantom{1}}$~symmetries are excluded.

\item Concerning proton decay, 
  if we wish to exclude upto dimension-five
  baryon-number violating operators, we are left with the DGSs: $R_3^{
  \phantom {1}}L_3^{\phantom{1}}$ ($\bsym{B}_3$), $R_3^2L_3^{\phantom{
  1}}$, $R_6^5L_6 ^2$ ($\bsym{P}_6$), $R_6^3L_6^2$ and $R_6^{\phantom
  {1}}L_6^2$. For $R_2^{\phantom {1} }$ ($\bsym{M}_p$), $R_3^{\phantom
  {1}}$ or $R_6^ {\phantom{1}}$, $QQQL$ and $\bar{U}\bar{U}\bar{D}\bar
  {E}$ must be suppressed by some mechanism due to the stringent
  bounds on proton decay, see {\it e.g.}
  Ref.~\cite{Harnik:2004yp,Larson:2004ji}. The DGS $L_3^{\phantom{1}}$
  is significantly constrained by the bounds on $\bar{
  U}\bar{D}\bar{D}$ from heavy nucleon decay \cite{Goity:1994dq}.
  
\item Now consider neutrino masses. Without right-handed neutrinos, we
  can generate masses at tree-level through the terms $LH_ uLH_u$ and
  $LH_u$ (via mixing with the neutralinos), or via loop diagrams
  involving $LL\bar{E}$ or $LQ \bar{D}$
  \cite{Grossman:1998py,Hirsch:2000ef,Davidson:2000uc,Abada:2002ju}.
  Hence, the DGSs $R_2^{\phantom{1}}$ ($\bsym{M}_p$),
  $R_3^{\phantom{1} }L_3^{\phantom{1}}$ ($\bsym{B}_3$) and
  $R_6^5L_6^2$ ($\bsym{P}_6$) can incorporate neutrino masses without
  right-handed neutrinos.\footnote{It is not possible to generate
  neutrino masses in the SSM in the case of $R_3$ or $R_6$. They allow
  for the lepton-number violating terms $QQQL$ and $\bar{U}\bar{U}\bar
  {D} \bar{E}$ but conserve $B-L$.}  However, right-handed neutrinos
  can easily be included as heavy Majorana fermions obeying
  Eq.~(\ref{majorana}). If the corresponding $U(1)_X$~charges allow
  Dirac neutrino mass terms, we obtain massive light neutrinos via the
  see-saw mechanism
  \cite{Minkowski:1977sc,Mohapatra:1979ia,see-saw,yanagida}. But in
  this case, $LH_uLH_u$ must be allowed by the $\maZ_N$~symmetry as
  well: invariance of the Dirac mass terms for neutrinos as well as
  the Majorana mass terms implies a $\maZ_N $~invariant $LH_uLH_u$
  term.

\end{itemize}

If we combine these phenomenological requirements, we are left with
only two DGSs: baryon-triality $\bsym{B}_3$, and proton-hexality
$\bsym{P}_6$. It is remarkable that these discrete symmetries also
survived in Sect.~\ref{rescaling}, \ie~they are discrete gauge
anomaly-free with {\it integer} heavy-fermion charges. However, we
would like to go a step further. In Sect.~\ref{intro}, we defined the
MSSM as the SSM restricted by $\bsym{M}_p$. When considering the MSSM
as a low-energy effective theory, the dangerous operators $QQQL$ and
$\bar{U}\bar{U}\bar{D}\bar{E}$ are {\it allowed}.  This is a highly
unpleasant feature of the MSSM. IR already pointed this out as an
advantage of the $R$-parity violating MSSM with $\bsym{B}_3$, which does
not suffer this problem. Here we propose a different solution: {\it We
define the MSSM as the SSM which is restricted by proton-hexality,
$\bsym{P} _6$}. The only phenomenological difference to the
conventional MSSM with $\bsym{M}_p$ is with respect to baryon-number
violation. However, given the stringent bounds on proton decay, we
find this new definition of the MSSM significantly better motivated. 
Note that in the language of IR, $\bsym{P} _6$ is a 
generalised matter-parity (GMP).

We conclude this section with some observations:
\begin{enumerate}
\item It is
interesting to note that, of the nine fundamental DGSs which allow the
$H_d H_u$ term, those with $N=6$ are each equivalent to the
requirement of imposing $R_2$ (\ie~matter-parity) {\it along with} one
of the four fundamental $\maZ_3$~symmetries. Explicitly one has
\beqn R_2^{\phantom{1}} \times R_3^{\phantom{1}}L_3^{\phantom{1}} &
\cong & R_6^5
L_6^2\,,\qquad \Longleftrightarrow \;~~~~ \bsym{M}_p \times \bsym{B}_3 
\,\cong\, \bsym{P}_6\label{cong1}\\
R_2^{\phantom{1}} \times R_3^{\phantom{1}} & \cong &
R_6^{\phantom{1}}\,,
\label{cong2}\\
R_2^{\phantom{1}} \times L_3^{\phantom{1}} & \cong & R_6^3L_6^2\,,
\label{cong3}\\
R_2^{\phantom{1}} \times R_3^2L_3^{\phantom{1}} & \cong & R_6^{\phantom{1}}
L_6^2\,.\label{cong4}
\eeqn
In the first line we have given the corresponding isomorphism in
terms of matter-parity, baryon-triality and proton-hexality. The
reason for this is that the Cartesian product of the cyclic groups
$\maZ_2$ and $\maZ_3$ is isomorphic to $\maZ_6$, {\it i.e.} $\maZ_2
\times \maZ_3 \cong \maZ_6$ \cite{Kurzweil:1977}. This becomes evident
by giving both possible isomorphisms $\maZ_2 \times \maZ_3 \rightarrow
\maZ_6$.
\beqn
(0,0)\mapsto 0,~~(0,1)\mapsto 2,~~(0,2)\mapsto 4,
~~(1,0)\mapsto 3,~~(1,1)\mapsto 5,~~(1,2)\mapsto 1,\label{iso1}~~&&\\
(0,0)\mapsto 0,~~(0,1)\mapsto 4,~~(0,2)\mapsto 2, ~~(1,0)\mapsto
3,~~(1,1)\mapsto 1,~~(1,2)\mapsto 5.\label{iso2}~~&& 
\eeqn 
As an example, we calculate the discrete charges in the case of
Eq.~(\ref{cong1}). Recalling the relations between $q_i$ and the
exponents $m$, $n$ and $p$ given in Eq.~(\ref{discharges}), we find
for the $\maZ_2 \times \maZ_3$ charges, where we compute modulo $N$
[{\it e.g.}  $q_{\bar{U}}=(-1,-1)=(1,2)$]:
\beqn
&&q_Q=(0,0),~~~~q_{\bar{U}}=(1,2),~~~~q_{\bar{D}}=(1,1),~~~~q_L=
(0,2),~~~~q_{\bar{E}}=(1,2),~~~~~~~~\notag\\
&&q_{H_d}=(1,2),~~~~q_{H_u}=(1,1),~~~
\label{z2z3}
\eeqn
and for the $\maZ_6 $ charges
\beqn
q_Q=0,~~~~q_{\bar{U}}=1,~~~~q_{\bar{D}}=5,~~~~q_L=4,~~~~
q_{\bar{E}}=1,~~~~q_{H_d}=1,~~~q_{H_u}=5.~~~\label{z6}
\eeqn
Both charge assignments are related by the isomorphism of
Eq.~(\ref{iso1}). Similarly, the $\maZ_2 \times \maZ_3$ and the
$\maZ_6 $ charges in Eqs.~(\ref{cong3}) and (\ref{cong4}) are related
by this isomorphism. In the case of Eq.~(\ref{cong2}) we have to apply
the isomorphism of Eq.~(\ref{iso2}).

\item In Ref.~\cite{Dreiner:2003yr}, a $U(1)_X$ gauge extended SSM was
  investigated, where all renormalisable MSSM superpotential terms
  have a total $X$-charge which is an integer multiple of $N$ [{\it
  cf.} Eq.~(\ref{xxtot})]. Then the conditions on the $U(1)_X$ charges
  were derived, in order to have a low-energy $\bsym{M}_p$ discrete
  symmetry. In Ref.~\cite{Dreiner:2006xw}, we derive the corresponding
  conditions for $\bsym{B}_3$ and $\bsym{P}_6$:

\begin{itemize}
\item $\bsym{M}_p$:~~ $3X_{Q_1}+X_{L_1}=2\cdot\mathbb{Z}$,\;\;\;
$X_{H_d}-X_{L_1}=2\cdot\mathbb{Z}-1$ 
\item $\bsym{B}_3$:~~ $3X_{Q_1}+X_{L_1}=3\cdot\mathbb{Z}\pm 1$, 
\;\;\;$X_{H_d}-X_{L_1}=3\cdot\mathbb{Z}$ 
\item $\bsym{P}_6$:~~ $3X_{Q_1}+X_{L_1}=6\cdot\mathbb{Z}\pm 2$, 
\;\;\;$X_{H_d}-X_{L_1}=6\cdot\mathbb{Z}- 3$
\end{itemize}

\item Next, we consider domain walls, which pose a severe cosmological
  problem if they occur \cite{Kibble:1976sj}. It is commonly held that
  a spontaneously broken discrete symmetry leads to domain walls. In
  particular, this is expected to occur in the SSM if the Higgs fields
  are charged under the $\bsym{Z}_N$ symmetry.  In contrast, we do not
  expect domain walls if the Higgs' discrete charges are zero.
  However, by this reasoning the first set of charges below
  Eq.~(\ref{rp}), ($q_{H_u}=1$, $q_{H_d}=-1$) implies the existence of
  domain walls, whereas the second set, standard matter-parity
  ($q_{H_u}=0$, $q_{H_d}=0$), does not. As stated in
  Sect.~\ref{linear-2}, these two symmetries are related by a simple
  hypercharge shift. They have the same low-energy superpotential and
  soft terms. Hence the resulting scalar potentials are identical
  apart from $D$-term contributions. Therefore the two theories have
  the same vacuum structure, and either both have or both do not have
  domain walls.
  
  If the SSM vacuum $\{\ups_{H_{d}},\ups_{H_{u}}\}$ has zero $\bsym
  {Z}_N$ charge, then it is unique. If it transforms non-trivially
  under $\bsym{Z}_N$ then there are upto $N$ distinct ground states $
  \{\ups_{H_{d}},\ups_{H_{u} }\}$, $\{{\ups_{H_{d}}}^\prime,{\ups_{H_
  {u}}}^\prime\}$, $\{{\ups_{H_{d}}}^{\prime\prime},{\ups _{H_{u}}
  }^{\prime\prime}\}$, ..., related by $\bsym {Z}_N$ transformations.
  In the latter case, there are however no domain walls, if the $
  \bsym{Z}_N$ transformation of the vacuum in a given domain can be
  compensated by a $U (1)_Y$ gauge transformation. Explicitly, we
  demand there exists a combined $\bsym{Z}_N\!+\!\bsym{Y}
  $-transformation $\bsym{T}$, such that $\bsym{T}(H_{d,u})=H_{d,u}$,
  \emph{i.e.}
\beq
\exists \,\alpha(x):\qquad \exp\left[\mbox{$\mathrm{i}\frac{2\pi}{N}$}
\cdot q_{H_{d,u}}+\mathrm{i}\alpha(x)\cdot Y_{H_{d,u}}\right] H_{d,u}
=H_{d,u},
\eeq
$\alpha(x)\in\mathbb{R}$ is the gauge parameter of $U(1)_Y$. This is
equivalent to
\beq 
\mbox{$\frac{2\pi}{N}$}\cdot q_{H_{d,u}}+\alpha(x)\cdot
Y_{H_{d,u}}=2\pi\cdot I_{d,u}\,,\qquad \mathrm{with}\;\;
I_{d,u}\in\mathbb{Z} \,.
\eeq
These two equations can be combined to
\beqn 
I_u
&=&\mbox{$\frac{1}{N\cdot Y_{H_d}}$}\cdot(q_{H_u}\cdot Y_{H_d} -
q_{H_d}\cdot Y_{H_u} + N\cdot Y_{H_u}\cdot I_d)\,,\\
\alpha(x) &=&\mbox{$\frac{2\pi}{N\cdot Y_{H_d}}$}\cdot (N\cdot
I_d-q_{H_d}).~~~ 
\eeqn 
The second equation defines the required gauge transformation. We can
simplify the first equation, using the hypercharge relation $Y_{H_u}
=-Y_{H_d}$
\beq
N\cdot (I_u+I_d) =q_{H_d}+q_{H_u}.  
\eeq 
This can only be fulfilled if the $\bsym{Z}_N$-charges of the
two Higgs, just like their hypercharges, are the inverse of each other (in
the sense of a mod~$N$ calculation).\footnote{If the two Higgs do not
have opposite $\bsym{Z}_N$-charges, the $\mu$-term is forbidden. This
then possibly enables PQ-invariance, which allows one to repeat the
argument above with $~\alpha(x)\cdot Y_{H_{d,u}}~$ replaced by
\mbox{$\alpha(x)\cdot Y_{H_{d,u}}+\beta\cdot P\!Q_{H_{d,u}}$}.} This
is equivalent to the requirement that the $\mu$-term is allowed by
$\bsym{Z}_N$. This is \emph{e.g.} the case for $\bsym{M}_p$
canonically, as the Higgs fields are uncharged: $(q_{H_d},q_{H_u})=
(0,0)$, $R_2$ $(1,1)$, $\bsym{B}_3$ $(2,1)$ and $\bsym{P}_6$
$(1,5)$. We stress that this argument does not rely on $U(1)_X$ being
non-anomalous ({\it cf.} Sect.~\ref{top-down}).

\end{enumerate}


\section{The Heavy-Fermion Sector}
\label{heavy}
\cleqn
An interesting question to ask is as follows: Given a DGS in Table~\ref{fundtab},
do I necessarily need heavy fermions in order to cancel the
anomalies? In the case of matter-parity, $R_2$, we can answer the
question by considering Eq.~(\ref{l3}). Here, the left-hand side
equals 3, while the right-hand side is
$~2\cdot\mathbb{Z}+\eta\cdot\mathbb{Z}$.  Recalling that the
$\eta$-term originates from heavy Majorana fermions [\textit{cf.}
Eq.~(\ref{ggx})], we find that the symmetry $R_2$ is only possible if
we include a heavy-fermion sector, \emph{e.g.} one right-handed
neutrino for each generation.

In the case of the other fundamental DGSs of Table~\ref{fundtab}, let
us assume the absence of heavy fermions in what follows. Under this
assumption, the anomaly cancellation conditions cannot be satisfied.
Inserting the discrete charges of Eq.~(\ref{discharges}) into
Eq.~(\ref{ggx}), we obtain
\beqn
13n+3p-3m\!\!&=&\!\! N \!\cdot \! \Bigg[2m_{H_d} + 2m_{H_u}+
\sum_k (6m_{Q_k} +3m_{\bar{U}_k} + 3m_{
\bar{D}_k} + 2 m_{L_k} +m_{\bar{E}_k})\Bigg] \!,~~~~~~
\label{hs3}
\eeqn
where $k$ is a generation index.  For even $N$, the right-hand side in
Eq.~(\ref{hs3}) is even. However, the left-hand side is odd for the
$\maZ_2$, $\maZ_6$ and $\maZ_{18}$ DGSs. Therefore heavy fermions are
necessary in these cases.

For the remaining $4+9$ $\maZ_{3}$ and $\maZ_{9}$ symmetries, the
right-hand side (RHS) of Eq.~(\ref{hs3}) can be both, even or odd. We
thus employ the cubic anomaly constraint of Eq.~(\ref{cub}). For the
$\maZ_{9}$~symmetries the RHS of Eq.~(\ref{cub}) is always a multiple
of $27$. The left-hand side (LHS) of the cubic anomaly condition,
given in Eq.~(\ref{LHS9}), is $-122\cdot 3 + 27\cdot\mathbb{Z}$, which
is not a multiple of $27$. Thus the fundamental $\maZ _9$~symmetries
also require heavy fermions.

For the four $\maZ_3$~symmetries the RHS of Eq.~(\ref{cub}) is always
a multiple of $9$. Eq.~(\ref{LHS3}) shows that the LHS of
Eq.~(\ref{cub}) is a multiple of $9$ only in the case of the
$R_3L_3$~symmetry. Hence the other three fundamental
$\maZ_3$~symmetries require heavy fermions.  But also $R_3L_3$ cannot
satisfy the anomaly constraints without a heavy-fermion
sector:\footnote{Here we disagree with Ib\'a\~nez's conclusion in
Ref.~\cite{Ibanez:1992ji}. See also Ref.~\cite{Hinchliffe:1992ad}.} 
Although $R_3L_3$ is neither ruled out by $\mathcal{A}_{GGX}=0$ nor
$\mathcal{A}_{XXX}=0$ alone, it is in conflict when combining the two
conditions; the LHS of Eq.~(\ref{cub}) for $R_3L_3$ yields $18$,
[\textit{cf.}  Eq.~(\ref{LHS3})], whereas the RHS is a multiple of 27,
as we now show. It is given by
\beq
 - \sum_i \left(3{q_i}^2m_iN + 3q_i{m_i}^2N^2 +{m_i}^3N^3   \right),
\eeq
where $i$ runs over all chiral superfields. The last two terms within
the parentheses are multiples of $27$, which is not true for the first
one.  However, evaluating the sum and applying our knowledge of the
$q_i$, we find
\beq
\sum_i 3{q_i}^2m_iN =  3N \cdot \Bigg[2 \cdot  m_{H_d}+ 2 \cdot  
m_{H_u} +  \sum_k \Big( 3\cdot 
m_{\bar{U}_k} +  3\cdot m_{\bar{D}_k} +   2\cdot m_{L_k} + 4\cdot 
m_{\bar{E}_k} \Big) 
\Bigg],\label{1.term}
\eeq
where $k$ denotes a generation index. The numerical coefficients
inside the brackets are the product of the squared discrete charges
and the multiplicity of the particle species. For example, we have 3
colours of quark fields $\bar{U}_k$ with $q_{\bar{U}_k} = -1$, thus
$3\cdot{q_{\bar{U}_k}}^2 = 3$. We can now adopt the $\mathrm{gravity}$-$ \mathrm{gravity} $-$U(1)_X$ 
anomaly constraint of Eq.~(\ref{hs3}) to
rewrite Eq.~(\ref{1.term}). Recalling that $N=3$, $n=0$ and $m=p=1$
for $R_3L_3$, we get
\beqn
\sum_i 3{q_i}^2m_iN &=&  -9 \cdot  \sum_k \Big( 6\cdot m_{Q_k} - 3\cdot 
m_{\bar{E}_k} \Big)\,, 
\eeqn
also a multiple of $27$. This completes our proof.

In conclusion: The 27 fundamental DGSs we have found are {\it only}
anomaly-free with a $U(1)_X $-charged heavy-fermion sector.


\section{A Top-Down Approach}
\label{top-down}
\cleqn

As outlined in Sect.~\ref{intro}, we have so far discussed a bottom-up
approach to DGSs. However, by definition, a DGS is inherently
connected to the anomaly structure of the underlying $U(1)_X$ gauge
theory. Here, we consider the DGSs from the latter perspective. We
investigate two topics in detail: {\it (i)} the definition of the DGSs
via the transformation of the superfields (superfield-wise) vs. the
definition via the transformation of the $\Gsm$ invariant operators
(operator-wise); {\it (ii)} the hypercharge shifts of
Eq.~(\ref{y-shift}).

At high energies, we start from a $\Gsm\times U(1)_X$ invariant
Lagrangian, with the $X$-charges scaled to be integers of minimal
absolute value. We leave it open at the moment whether $U(1) _X$ is
anomalous or not. Below $M_X$, $U(1)_X$ is assumed to be broken by a
single left-chiral flavon superfield $\Phi$ (or by two left-chiral
superfields $\Phi,\Phi'$ with opposite $X$-charges, see
Sect.~\ref{gauged}), which is uncharged under $\Gsm$. If in our model
\emph{e.g.} the operator $L_iL_j\bar{E}_k$ is not $U(1)_X $-invariant,
then the non-renormalisable superpotential\footnote{The following
arguments in this Sect.~proceed analogously for the K\"ahler
potential.}  operator
\beq 
\Phi^{\!~-~\frac{X_{L_i}+X_{L_j}+X_{{\bar {E}}_k}}{X_{\Phi}}}~
\times~ L_iL_j\bar{E}_k
\eeq
is. However, due to the cluster decomposition principle (CDP)
\cite{Wichmann:1963}, the Lagrangian exhibits only non-negative
integer exponents of the fields
\cite{Weinberg:1995mt,Weinberg:1996kw}. Therefore the above term is
forbidden if $\,\frac{X_{L _i} +X_{L_j}+X_{\bar{E}_k}}{X_{\Phi}}\,$ is
fractional. After $U(1)_X $-breaking, the operator $L_iL_j\bar{E}_k$
is not generated, since its non-renormalisable ``parent term" is
non-existent. Therefore the constraints of the CDP persist. Whether an
operator is allowed or not in the low-energy Lagrangian boils down to
whether its overall $X$-charge is an integer multiple of $X_{\Phi}$.
Thus at low energy, we decompose the $X$-charges as in
Eq.~(\ref{charges}) and the remaining DGS under which the
\emph{superfields} transform is a $\maZ_{|X_{\Phi}| }$.

Next consider the operators in the superpotential. Analogous to
Eq.~(\ref{charges}), the overall $X$-charge, $X_{\mathrm{total}}$, of
any $\Gsm$-invariant product of MSSM chiral superfields satisfies
\beq
X_{\mathrm{total}}=q_{\mathrm{total}}+m_{\mathrm{total}}\cdot|X_{\Phi}|\,,
\qquad \mathrm{with} 
\quad q_{\mathrm{total}}\equiv \sum q_i\,, \quad m_{\mathrm{total}}
\equiv\sum m_i\,.  
\label{X-tot}
\eeq
If a certain operator is forbidden by the CDP, then the $|X_{\Phi}|
^{th}$ power of this term has $q_{\Mrm{total}}=0\,\mathrm{mod}
(|X_\Phi|)$. However, the superpotential operators are further
restricted by $\Gsm$. Therefore the $\bsym{Z}_{|X_\Phi|}$-charges are
possibly such that a power smaller than $|X_{\Phi}|$ suf\-fices to get
$q_{\Mrm {total}}=0\,\mathrm{mod} (|X_\Phi|)$, for {\it all}
superpotential operator. As an example, suppose $|X_{\Phi}|=24 $ and
the superfields obey a $\bsym{Z}_{24}$. Due to $\Gsm$, it may very
well be that for {\it all} operators $q_{\mathrm{total }}$ is even.
Operator-wise we then have a $\bsym{Z}_{12}$ instead of a $\bsym{Z}
_{24}$.  Furthermore, we can integrate out the heavy particles below
their mass scale. When considering only the superfields of the SSM
their respective $q$'s could \emph{e.g.} be only multiples of~3. The
SSM superfields alone then obey a $\bsym {Z}_{24/3}=\bsym{Z} _{8}$
symmetry (\emph{cf.} Sect.~\ref{rescaling}) and the SSM
superfield-wise $\bsym{Z}_{8}$ constitutes an SSM-operator-wise
$\bsym{Z}_{4}$.

We now consider a generation-independent $U(1)_X$ extension of the
SSM, which is the high-energy origin of the DGS. We include
right-handed neutrinos, $\bar N_i$. We demand that for the $U(1) _X$
charge assignments: {\it(i)} the Yukawa mass terms $QH_d\bar{D}$,
$QH_u\bar{U} $, $LH_d\bar{E}$, and $LH_u{\bar N}$ are invariant, and
{\it (ii)} the anomalies $\mc{A} _{CCY}$, $\mc{A}_{WW Y}$,
$\mc{A}_{GGY }$, $\mc{A}_ {CCX}$, $\mc{ A}_{WWX}$, $\mc{A}_{GG X} $,
$\mc{A}_{YYY }$, $\mc{A}_ {YYX}$, $\mc{A}_ {YXX}$ and $\mc{A} _{XXX}$
all vanish.  We can then express the $X$-charges in terms of two
unknowns
\beqn
\begin{array}{rclrclrcl}
X_{\bar{D}}\!\! &=& \!\!\!\!-X_Q-X_{H_d},&\ \ X_{\bar{U}}\!\!&=& \!\!\!\!-X_Q+X_{H_d},
&X_{L}\!\!&=& \!\!\!\!-3X_Q, \\
X_{\bar{E}}\!\!\! &=& \!\!\!\!3X_Q-X_{H_d},&
X_{\bar{N}}\!\! &=& \!\!\!\!3X_Q+X_{H_d},&X_{H_u}\!\!&=& \!\!\!\!-X_{H_d}\,.
\end{array}
\label{BBBBB} 
\eeqn 
Furthermore, we obtain the well known result that $U(1)_X$ is
necessarily a linear combination of $U(1)_Y$, \ie\ hypercharge, and
$U(1)_{\mathrm{B-L}}$ (see for example
Ref.~\cite{Weinberg:1996kr,Mohapatra:1998rq,Chamseddine:1995rs}) 
\beq
X_i=\frac{X_i^{\mathrm{B-L}}}{X_Q^{\mathrm{B-L}}}\cdot C_1+\frac{Y_i}
{{Y}_Q}\cdot C_2\,,
\label{general-U1}
\eeq
where $C_{1,2}$ are free real parameters, such that the $X$-charges
are integers, as was required earlier. Eq.~(\ref{BBBBB}) can then be
reexpressed in terms of $C_{1,2}$
\beqn 
\begin{array}{rclrclrcl}
X_Q&=&C_1+C_2\,,& X_{\bar{D}}&=&-C_1+2C_2\,,& X_{\bar{U}}&=&-
C_1-4C_2\,, \\
X_L&=&-3C_1-3C_2\,,&  X_{\bar{E}}&=&3C_1+6C_2\,,&
X_{\bar{N}}&=&3C_1\,,\\
X_{H_d}&=&-3C_2\,,& X_{H_u}&=&3C_2\,.& &&
\end{array}
\label{X-c1-c2}
\eeqn
For $2C_1=-5C_2 $, we obtain a theory with $SU(5)$ invariant
$X$-charges. For $C_1\not=0$ the right-handed neutrinos are charged
and the see-saw mass term $\bar N_i\bar N_j$ is forbidden. And of
course for $C_2=0$ we obtain $U(1)_{\mathrm{B-L}}$.

At low-energy, we performed the hypercharge shift of the DGS,
Eq.~(\ref{y-shift}). As we argued, this hypercharge shift is
irrelevant for the structure of the low-energy superpotentials. From
the top-down approach, however, a different choice of $C_2$
corresponds to a hypercharge shift of the SSM $X$-charges, which in
turn corresponds to a hypercharge shift of the corresponding
$\bsym{Z}_N$. How does this change the high-energy theory?  The gauge
boson and fermionic kinetic terms in the Lagrangian are
\beq
\scr{L}=-\frac{1}{4}F_X^2-\frac{1}{4}F_Y^2+\sum \overline\psi_k
\left(i\partial_\mu -g_XX_k A_\mu^X-g_YY_k A^Y_\mu\right)\gamma^\mu\psi_k\,.
\label{eich}
\eeq
Here $F_{X,Y}^2$ are the squared field strength tensors, and $A_\mu^
{X,Y}$ are the corresponding gauge potentials. We see that a
simultaneous orthogonal rotation in the fields $(A_\mu^X,A_\mu^Y)$ and
the charges $(g_X X_k,g_Y Y_k)$ leaves the Lagrangian unchanged. But
different choices of $C_2$ in Eq.~(\ref{general-U1}), which correspond
to hypercharge shifted (not rotated) theories, lead to distinct gauge
theories in Eq.~(\ref{eich}). They differ in their $X$-charges and
thus in their scattering cross sections. They are therefore, in
principle, experimentally distinguishable at energies $\sqrt{s}={\cal
O}(M_X)$.  However, at the LHC, we can only determine the low-energy
DGS. We can {\it not} determine $C_2$ of Eq.~(\ref{general-U1}). When
attempting to interpret the LHC results in terms of an underlying
unified theory it is important to keep this ambiguity in mind.

Let us now focus on the $\Phi\!+$SSM-sector, \ie~including the flavon
field(s). Using the methods of
Refs.~\cite{Dreiner:2003yr,Dreiner:2006xw}, we can compute the total
$X$-charge of \emph{any} $\Gsm$-invariant superpotential term and
obtain
\beq\label{xxtot}
X_{\mathrm{total}}^{\mathrm{SSM}}=\mathbb{Z}\cdot(3X_{Q_1}+X_{L_1})
+\mathbb{Z}\cdot(X_{H_d}-X_{L_1})+\mathbb{Z}\cdot(X_{H_d}+X_{H_u})+
\mathbb{Z}\cdot|X_\Phi|\,,
\eeq
where $\mathbb{Z}$ again denote arbitrary and independent integers.
Using Eq.~(\ref{charges}), this gives
\beq\label{qqtot}
q_{\mathrm{total}}^{\mathrm{SSM}}=\mathbb{Z}\cdot(3q_{Q}+q_{L})+
\mathbb{Z}\cdot(q_{H_d}-q_{L})+\mathbb{Z}\cdot(q_{H_d}+q_{H_u})+
\mathbb{Z}\cdot|X_\Phi|\,.
\eeq 

We have seen that a hypercharge shift of the $X$-charges leads to a
\emph{new} $U(1)_X$ gauge theory. Such a shift is however only
possible for an originally anomaly-free model (see \emph{e.g.} the
completely fixed $X$-charges in Ref.~\cite{Dreiner:2003yr}) and yields
an alternate anomaly-free model.  Plugging the $X$-charges of
Eq.~(\ref{general-U1}) into Eq.~(\ref{xxtot}), we find
\beq
X_{\mathrm{total}}^{\mathrm{SSM}}=\mathbb{Z}\cdot3C_1+\mathbb{Z}
\cdot|X_\Phi|,
\eeq 
of course independent of $C_2$ and thus of hypercharge. So all the
results on the operator-wise DGS coming from $U(1)_X$ are solely
determined by $C_1$ and $|X_\Phi|$. This characteristic, which we
demonstrated for a simple example, also holds for all non-anomalous
models. This is why we could shift away $r$ in Sect.~\ref{linear}. For
$C_1=-C_2$, {\it i.e.} $X_Q=0$, the field-wise and operator-wise definition of
the DGS coincide.

Equipped with the $X$-charges in Eq.~(\ref{general-U1}), we now
demonstrate in two examples the emergence of distinct operator- and
superfield-wise DGSs from the $U(1)_X$.

\begin{itemize}
  
\item {$C_1=1,\,C_2=0,$} supplemented by a vector-like pair of
flavon superfields, $X_{\Phi}=6$, $X_{\Phi^\prime}=-6$. Hence the
Yukawa operators have the total $X$-charge $X_{LH_{d} {\bar E}}=X_{Q
H_{d} {\bar D}}=X_{Q H_u{\bar U}} =X_{LH_u\bar{N}}=X_{H_{d} H_{u}}=0,$
but $X_{LL{\bar E}}=X_{LQ{\bar D}}=X_{{\bar U}{\bar D}{\bar D}}=X_{LH_
{\!u}} = -3$.

To have \emph{e.g.}~$LL{\bar E}$ generated after $U(1)_X$ breaking
would require $\sqrt{\Phi~}\cdot LL{\bar E}\,$, which is not allowed
due to the CDP. With Eq.~(\ref{charges}) we get a superfield-wise
$\maZ_{6}$, with $q_{Q}=1$, $q_{\bar{D}}=q_{\bar{U}}= 5$, $q_{L}=q_{
\bar{E}}=q_{\bar{N}}=3$, ${q_{H_d}}=q_{H_{u}}=0$.  Plugging these 
into Eq.~(\ref{qqtot}), one finds that any superpotential term has an
overall $q$-charge which is an integer multiple of either 3 or 6. Thus
the actual DGS of the \emph{operators} is a $\bsym{Z}_{\frac{6}{3}}=
\bsym{Z}_{2}$ symmetry. This is matter-parity, in fact.

\item $C_1=2$, $C_2=1$ results in $X_{Q}=3$, $X_{\bar{D}}=0$, $X_{
    \bar{U}}= -6$, $X_{L}=-9$, $X_{\bar{E}}=12$, $X_{\bar{N} }=6 $,
    $X_{ H _{d}}=-3$, $X_{H_{u}} =3$, again supplemented by $X_{
    \Phi}=6$, $X_{\Phi^\prime}=-6$. This leads to
    $q_{Q}\!=\!q_{L}\!=\!{q_ {H_d}}\!=\!q_{H _{u}}\!=\!3$, $q_{
    \bar{D}}\!=\!q_{\bar{U}}\!= \!q_{\bar{E}}\!=\!q_{\bar{N}}\!=\!0
    $. The DGS appears to be a $\bsym{Z}_{\frac{6}{3}}=\bsym{Z}_2
    $. However, inserting the charges into Eq.~(\ref{qqtot}), we find
    no DGS whatsoever.
\end{itemize}

Another example, more elaborate and flavour-dependent, is the fourth
model in Table~2 in Ref.~\cite{Dudas:1995yu}. It does not cause any
DGS after $U(1)_X$ breaking, as our second example. The prefactors of
the free parameter $q$ (their notation!) are nothing but the usual
hypercharges.

The argument that a superfield-wise $\maZ_{|X_{\Phi}|}$ causes an
operator-wise $\maZ_{|X_{\Phi}|/N}$ is independent of whether the
$U(1)_X$ has anomalies which are cancelled via Green-Schwarz
\cite{Green:1984sg} or whether the $U(1)_X$ is non-anomalous.  The
anomalous $X$-charges given in Table~7, Ref.~\cite{Dreiner:2003yr},
display a SSM superfield-wise $\bsym{Z}_{300}$ symmetry, but operator-wise 
constitute a $\bsym{Z}_{2}$, as can be seen by plugging the
corresponding discrete charges into Eq.~(\ref{qqtot}). \emph{A priori}
it is hence not clear whether, \emph{e.g.}, a superfield-wise $\bsym
{Z}_{300}$ gives rise to an operator-wise $\bsym{Z}_{300}$, $\bsym{Z}
_{150}$, $\bsym{Z}_{100}$,..., $\bsym{Z}_{2}$ or even $\bsym {Z}_{1}$
(trivial).

In summary, from a top-down point of view hypercharge shifted theories
are not equivalent. They are, in principle, experimentally
distinguishable by high-energy scattering experiments. If they are
anomaly-free, they lead to equivalent low-energy discrete gauge
theories and are not distinguishable at the LHC. But even a
non-anomalous and an anomalous set of $X$-charges are equivalent from
the low-energy point of view if they lead to the same operator-wise
DGS.


\section{A Gauged $\bsym{P_6}$ Model}
\label{gauged}
\cleqn
In this section, we explicitly present a generation-dependent 
$U(1)_X$~gauge model, constructed in collaboration with C.~A.~Savoy 
and S.~Lavignac. $U(1)_X$ is spontaneously broken to proton-hexality, 
$\bsym{P}_6$. We consider this a demonstration of existence, not 
necessarily an optimised model. Concerning the origin of the needed 
non-renormalisable interaction terms, there are several sources 
imaginable (see, {\it e.g.}, \cite{Ibanez:1994ig}): Either the terms occur 
near the string scale or they are generated by integrating out heavy 
vector-like pairs of $G_{\mathrm{SM}}$ charged states (the so-called 
Froggatt-Nielsen mechanism \cite{Froggatt:1978nt}). Here we adopt the 
first viewpoint and thus use a simple operator analysis. 
We assume the $U(1)_X$ breaking superfields to be 
suppressed by $M_{\mathrm{grav}}$, \emph{e.g.} $Q_1H_u\bar{U}_1$
derives from  $(\Phi_+/M_{\mathrm{grav}})^8\cdot Q_1H_u\bar{U}_1$.

We first list in Table~\ref{Table10000} the $U(1)_X$ charges of all the 
chiral superfields in our model. 
\begin{table}
\begin{center}
\begin{tabular}{|c|}
\hline $\phantom{\Bigg|}X_{\Phi_+} =6,~~~~$  $X_{\Phi_-}=-6$  \\
\hline
\end{tabular}~~~~~~
\begin{tabular}{|c|}
\hline $\phantom{\Bigg|}X_{H_d}=1,~~~~$  $X_{H_u}=-49$  \\
\hline
\end{tabular}
\end{center}
\begin{center}
\begin{tabular}{|c|c|c|c|c|c|}
\hline
\bf{Generation}\phantom{$\Big|$} $\bsym{i}$~ &~~~ $
\bsym{X_{Q_i}}$~~~ 
 &~~~ $\bsym{X_{\bar{U}_i}}$~~~ &~~~ $\bsym{X_{\bar{D}_i}}$ ~~~
&~~~ $\bsym{X_{L_i}}$
~~~ &~~~ $\bsym{X_{\bar{E}_i}} $~~~       \\\hline
1\phantom{\Bigg|} & $-12$ &$13$ &
$-25$ &$40 $ &$-77 $
  \\  \hline
2\phantom{\Bigg|} & $-12$ &$37$ &
$-13$ &$40 $ &$-17 $
  \\  \hline

3\phantom{\Bigg|} & 0 &$49$ &
$-13$ &$40 $ &$-53$
  \\  \hline
\end{tabular}\end{center}
\begin{center}
\begin{tabular}{|c|}\hline$\phantom{\Bigg|}X_{A_{\mathrm{D1}}} =-\frac{27}{2},~~~X_{A_{\mathrm{D2}}}=-\frac{45}{2},~~~X_{A'_{\mathrm{D1}}}=\frac{1}{2},~~~X_{A'_{\mathrm{D2}}}=\frac{71}{2},~~~X_{A_{\mathrm{M}}}=3$\\\hline\end{tabular}
\end{center}
\caption{The $U(1)_X$ charges of all chiral superfields in our model. 
$\Phi_\pm$ break $U(1)_X$, the $A_{...}$ are $G_{\mathrm{SM}}$ 
uncharged heavy particles.}
\label{Table10000}
\end{table}
The $G_{\mathrm{SM}}$ singlets $\Phi_\pm$ constitute the vector-like 
pair of $U(1)_X$ breaking superfields with equal VEVs. 
The $A_{...}$ are $G_{\mathrm{SM}}$ 
singlets as well but do not aquire VEVs, we introduce them solely for the 
sake to cancel $\mathcal{A}_{GGX}$ and $\mathcal{A}_{XXX}$. All the other
 (mixed) anomalies vanish within the particle content of the SSM.

The breaking of $U(1)_X$ generates the MSSM Yukawa coupling constants 
with textures that produce the observed fermionic mass spectrum 
as well as acceptable mixing matrices. Furthermore, $U(1)_X$ leaves a 
$\boldsymbol{Z}_{12}$ symmetry as a remnant which, after integrating out the 
$A_{...}$, yields $\boldsymbol{P}_6$:
\begin{itemize}
\item With 
\beq
\epsilon \equiv \frac{\langle \Phi_{\pm}\rangle}{M_{\mathrm{grav}}} = 0.22,
\eeq
we obtain an effective superpotential which contains the first line of 
Eq.~(\ref{superpot}) and the mass terms for the left-handed 
neutrinos ($h^{\nu}_{ij}/M_\nu \cdot L_i H_u L_j H_u$), where
\beq
\bsym{h^U} \sim 
 \begin{pmatrix} 
   \epsilon^8  & \epsilon^4  & \epsilon^2  \\  
   \epsilon^8  & \epsilon^4  & \epsilon^2  \\  
   \epsilon^6  & \epsilon^2  & 1  \\  
 \end{pmatrix},~~~
\bsym{h^D} \sim \epsilon^2 \cdot
 \begin{pmatrix} 
   \epsilon^4  & \epsilon^2  & \epsilon^2  \\  
   \epsilon^4  & \epsilon^2 & \epsilon^2  \\  
   \epsilon^2  & 1  & 1  \\  
 \end{pmatrix},~~~
\bsym{h^E} \sim \epsilon^2 \cdot
 \begin{pmatrix} 
   \epsilon^4  & \epsilon^2  & 1  \\  
   \epsilon^4  & \epsilon^2  & 1  \\  
   \epsilon^4  & \epsilon^2  & 1  \\  
 \end{pmatrix},
\eeq
\beq
\mu \sim \epsilon^8 \cdot M_\mu,~~~~~~~
\bsym{h^{\nu}} \sim \epsilon^3 \cdot 
\begin{pmatrix} 1 & 1 & 1 \\  1 & 1 & 1 \\ 1 & 1 & 1 \end{pmatrix}.
\eeq
To get the $\mu$ term and the neutrino masses of the correct order of 
magnitude, we rely on the existence of intermediate mass scales: 
$M_\mu \sim 10^{8}~\mathrm{GeV}$ (which's necessity has been already 
anticipated by Refs.~\cite{Dudas:1995yu,Jack:2003pb} for 
anomaly-free Froggatt-Nielsen models without heavy $G_{\mathrm{SM}}$ 
charged matter) and $M_\nu \sim 10^{12}~\mathrm{GeV}$.
After diagonalisation one gets for the masses of the electrically charged 
SM fermions  $m_u:m_c:m_t\sim\epsilon^8:\epsilon^4:1$,
$m_d:m_s:m_b\sim\epsilon^4:\epsilon^2:1$, 
$m_e:m_\mu:m_\tau\sim\epsilon^4:\epsilon^2:1$, 
$m_\tau: m_b:m_t\sim \epsilon^2 :\epsilon^2 : 1$. For the mixing matrices 
we get an anarchical MNS matrix, which is compatible with experiment, 
see {\it e.g.} Refs.~\cite{Hall:1999sn,Haba:2000be,deGouvea:2003xe}, 
as well as a CKM matrix which looks like
\beq
\bsym{V^{\mathrm{CKM}}}\sim 
 \begin{pmatrix} 
   1  & 1  & \epsilon^2  \\  
   1  & 1  & \epsilon^2  \\  
   \epsilon^2  & \epsilon^2  & 1  \\  
 \end{pmatrix}.~~~
\eeq
Thus we have to rely on some moderate fine-tuning  among the unknown 
$\mathcal{O}(1)$ coefficients to be entirely satisfactory.\\
Furthermore, we get the following mass terms for the heavy fields:
\beq
\epsilon^6 \cdot M_{\mathrm{grav}}~A_{\mathrm{D1}} A_{\mathrm{D2}},~~~~~
\epsilon^6 \cdot M_{\mathrm{grav}}~A'_{\mathrm{D1}} A'_{\mathrm{D2}},~~~~~
\epsilon \cdot M_{\mathrm{grav}}~A_{\mathrm{M}} A_{\mathrm{M}}.
\eeq
\item After $U(1)_X$ breaking we are left with an overall $\bsym{Z}_{12}$ DGS, 
since  $|X_{\Phi_{\pm}}|=6$ and all SSM particles' $X$-charges  are 
integers and the  $A_{...}$'s  $X$-charges  half-odd integers.
But as can be seen above, the $A_{...}$ are quite heavy, so that 
they all can be integrated  out at around 
$\epsilon^6M_{\mathrm{grav}}\sim 10^{14}~\mathrm{GeV}$, 
leaving the  \emph{fundamental} (in the sense of 
Section~\ref{rescaling})  DGS $\bsym{P}_{6}$.

\end{itemize}


\section{Summary}
\label{summary}
\cleqn
In summary, we have systematically investigated discrete gauge
symmetries $\bsym{Z}_N$, for arbitrary values of $N$. We have
classified the anomaly-free theories, depending on whether the
necessary (see Sect.~\ref{heavy}) heavy fermions are restricted to
integer $X$-charges or not. Through a rescaling of the $X$-charges, we
have for a low-energy point of view reduced this infinite set to a
finite fundamental set: All theories related by rescaling lead to the
same low-energy superpotential. For this fundamental set we have
investigated the phenomenological properties in detail. We have found
two outstanding DGSs, the second of them being beyond IR: \emph{(i)}
baryon-triality, $\bsym{B}_3$, which allows for low-energy
lepton-number violation, but no dimension-five or lower proton decay
operators, and \emph{(ii)} proton-hexality, $\bsym{P}_6$. The latter
has a renormalisable superpotential which conserves lepton- and
baryon-number and prohibits non-renormalisable dimension-five proton
decay operators. This is one of the main results of this paper and we
propose $\bsym{P}_6$ as {\it the} new discrete gauge symmetry of the
MSSM, instead of matter-parity. Both baryon-triality and
proton-hexality are free of domain walls.


\vspace{0.5cm}

\section*{Acknowledgments}
We are grateful to Graham G. Ross, Michael Flohr, Howie Haber, Michael
Dine, St\'ephane Lavignac, Borut Bajc, and especially Carlos A. Savoy
for helpful discussions, correspondence, or comments. C.L. thanks the
SPhT at the CEA-Saclay and M.T. the Physikalisches Institut, Bonn,  
for hospitality. M.T. greatly appreciates funding by a Feodor Lynen 
fellowship of the Alexander-von-Humboldt-Foundation.  This project 
is partially supported by the RTN European Program No. MRTN-CT-2004-503369.

\vspace{0.5cm}


\begin{appendix}

\section{The Cubic Anomaly}
\label{app-cubic}
\cleqn

In this appendix, we explicitly derive Table~\ref{cubtab}. We thus
restrict ourselves to integer charges for {\it all} chiral superfields
\cite{Ibanez:1991hv,Ibanez:1991pr} and investigate the
resulting consequences of the cubic anomaly constraint on possible
DGSs. Using Eq.~(\ref{discharges}), we can express the left-hand side
(LHS) of Eq.~(\ref{cub}) in terms of $n$, $p$ and $m$
\beqn
\mathrm{LHS} &=& -\, n\cdot\left(13n^2+18np -21nm+18p^2+21m^2 \right) 
 \notag\\&& + \;p \cdot\left( -3p^2+9pm+9m^2 \right) +3m^3\,,\label{LHS}
\eeqn
where we have made use of the fact that there are only 3 generations
in the SSM.  Even when disregarding the restrictions on the
heavy-particle content arising from the linear constraints, the
right-hand side (RHS) of Eq.~(\ref{cub}) can \textit{not} take on
arbitrary integer values. We shall denote it as ${\rm RHS}\equiv{\rm
  RHS}_1+{\rm RHS}_2+{\rm RHS}_3$, with a term for each line in
Eq.~(\ref{cub}). We now investigate these terms individually.

\begin{enumerate}
  
\item[(a)] \underline{${\rm RHS}_2$}: Factoring $N$, we see that the 
term ${\rm RHS}_2$ contributes a multiple of $N$ to the RHS. However,
it can not necessarily take on every possible multiple of $N$,
regardless of what the choice of heavy particles is.  For $(3\,|N)$,
we can again write $N=3N'$ ($N'\in\mathbb{N}$), and rewrite the last
term as $p_j^3N^3=3p_j^3 N^2N'$. We can thus factor $3N$ and therefore
the term ${\rm RHS}_2$ can take on \textit {at most} values $\in$
$3N\cdot\mathbb{Z}$. By adding appropriate sets of heavy Dirac
particles with simple charges, it is straightforward to show that {\it
any} multiple of $3N$ can be obtained. For DGSs with $\neg \,(3\,|N)$,
any element $\in N\cdot\mathbb {Z}$ can be obtained.
 
\item[(b)] \underline{${\rm RHS}_3$}: For odd $N$, $p_{j^\prime}^{\prime
}$ has to be even [see Eq.~(\ref{majorana})], so that the term ${\rm
RHS}_3$ is an element of $N^3\cdot \mathbb{Z}$. For even $N$, ${\rm
RHS}_3$ can take on all values $\in$ $\left(\frac{N}{2 }\right)^3\cdot
\mathbb{Z}$.
  
\item[(c)] \underline{${\rm RHS}_1$}: The first two terms in ${\rm
RHS}_1$ are multiples of $3N$, which is included in (a), above.
Similarly, the third term is a multiples of $N^3$ and therefore
already included in (b).

\end{enumerate}
Summarising, the RHS of Eq.~(\ref{cub}) can only take on values
obeying
\beqn
\mathrm{RHS} &=& 3N\cdot \mathbb{Z} + \left\{ \begin{array}{ll} N^3 \cdot 
\mathbb{Z}, & \quad N=\mathrm{odd},\\  \left(\frac{N}{2}\right)^3 \cdot 
\mathbb{Z}, & \quad N=\mathrm{even}, \end{array}   \right.~~~~~\mathrm{for}~~~
(3\,|N)\,, \\
&=& \left\{ \begin{array}{lll} 
\left.  \begin{array}{l} 9N' \cdot \mathbb{Z}, \end{array}\right. & 
\left.  \begin{array}{r}~~~~(3\,|N), \end{array}\right. & 
\quad N=\mathrm{odd},\\[0.1cm]  
\left.  \begin{array}{l} 9\frac{N'}{2} \cdot \mathbb{Z}, \\  9N' \cdot 
\mathbb{Z},\end{array}\right. &  
\left.  \begin{array}{r} \neg\,(12\,|N),\\ (12\,|N),  \end{array}\right\} & 
\quad N=\mathrm{even}, 
\end{array}   \right. \label{RHS}
\eeqn
where $N'=N/3$, as before. Furthermore
\beqn \mathrm{RHS} &=& N\cdot \mathbb{Z} + \left(\frac{N}{2}\right)^3
\cdot \mathbb{Z},
\qquad N=\mathrm{even},\qquad\mathrm{for}\quad\neg\,(3\,|N).\label{RHS2}
\eeqn

Now consider the LHS, while taking the linear constraints of
Sect.~\ref{linear} into account. Again, we investigate the cases $\neg
\,(3\,|N)$ and $(3\,|N)$ separately.
\begin{enumerate}
\item {\bf $\bsym{\neg \, (3\,|N)}$:} The DGSs of
  Eq.~(\ref{rp}), satisfying the linear constraints, require $n=p=0$
  and $m=\frac{N}{2}$. Thus the LHS becomes $3\cdot\left(\frac{N}{2}
  \right )^3$ [\textit{cf.} Eq.~(\ref{LHS})]. Comparing with
  Eq.~(\ref{RHS2}), we see that the cubic anomaly cancellation
  condition can be satisfied for all anomaly-free DGSs of
  Table~\ref{lintab} with $\neg\,(3\,|N)$, \textit{i.e.} the cubic
  anomaly results in no new constraint.
  
\item {\bf $\bsym{(3\,|N)}$:} We consider the remaining four 
categories of Table~\ref{lintab} in turn.

\begin{itemize}
  
\item[$(i)$] {$\bsym{(3\,|N)}$, $\bsym{N=\mathrm{odd}\!:}$}
Eq.~(\ref{RHS}) shows that the RHS must be a multiple of $9N'$.
Therefore the LHS must also be a multiple of $9N'$. From the
corresponding row in Table~\ref{lintab}, we see that in this case
$n=0$, $p=\ell_p N'$ and $m=\ell_m N'$. Inserting this into the LHS
as given in Eq.~(\ref{LHS}) yields
  \beqn 
   \mathrm{LHS}&=&\left(-3{\ell_p}^3 +9{\ell_p}^2\ell_m +
    9\ell_p{\ell_m}^2 + 3{\ell_m}^3\right) \cdot {N'}^3\,.\label{LHS3}
  \eeqn
For the case where $\ell_p=\ell_m$, we can satisfy the condition
$(9N'\,|\mathrm{LHS})$ for all $N$, which are subsumed in this
category, \textit{i.e.} any $N\in 6\cdot \mathbb{N}+3$. The remaining
cases of Table~\ref{lintab}, where $\ell_p\not=\ell_m$, require
$(3\,|{N'}^2)$, and hence $N=18\cdot \mathbb{N} + 9$.

\item[$(ii)$] {$\bsym{(3\,|N),\; N=\mathrm{even}\!:}$} From 
Table~\ref{lintab} we have in this case: $n=0$, $p=\ell_p N'$ and
$m=s_m \frac{N'}{2}$. The LHS then becomes
\beqn
\mathrm{LHS}&=&\left(-24{\ell_p}^3 +36 {\ell_p}^2s_m + 
18\ell_p{s_m}^2 + 3{s_m}^3\right)\cdot \left(\frac{N'}{2}\right)^3.
\eeqn
Due to the form of the RHS for $\neg\,(12\,|N)$ [\textit{cf.} 
Eq.~(\ref{RHS})] , we need $(9\frac{N'}{2}\, |\mathrm{LHS})$. This
leads to three non-trivial possibilities for arbitrary $N$ in this
category ($N=12\cdot\mathbb{N} + 6$): $[\ell_p=0 \wedge s_m=3]$, $[\ell
_p=1 \wedge s_m=2]$ and $[\ell_p=1 \wedge s_m=5]$. All DGSs can
satisfy the cubic anomaly constraint if $(3\,|{N'}^2)$, hence if $
N=36\cdot\mathbb{N} +18$.

\noindent Considering $(12\,|N)$ yields exactly the same three sets $(\ell_p,
s_m)$ for non-trivial possible DGSs with arbitrary $N \in
12\cdot \mathbb{N}$. All DGSs are allowed if $(3|{N'}^2)$, {\it
  i.e.}  for $N=36\cdot\mathbb{N}$. 

\noindent Combining the results for
$\neg\,(12\,|N)$ and $(12\,|N)$, we find that for each $N \in 6\cdot
\mathbb{N}$ there are three allowed non-trivial DGSs. Taking $N
\in 18 \cdot \mathbb{N}$, \textit{any} DGS  
satisfying the linear constraints is compatible with the cubic
constraint.

\item[$(iii)$] $\bsym{(9\,|N),\; N=\mathrm{odd}\!:}$ From
  Table~\ref{lintab} we obtain in this case $n=N'$, $p=(1+3\ell_p)
  N''$ and $m=(2+3\ell_m)N''$. Inserting this into Eq.~(\ref{LHS})
  gives
\beqn 
\mathrm{LHS}&=&\Big[ -27 {\ell_p}^3 + 27{\ell_p}^2 (-5+3\ell_m) + 9 \ell_p
(-23 +18 \ell_m+ 9{\ell_m}^2) \notag \\ 
&& \quad+ (-122+18\ell_m-108{\ell_m}^2+27{\ell_m}^3 ) \Big]\cdot N' 
\cdot {N''}^2.\label{LHS9}
\eeqn 
As $122$ is not a multiple of $9$, whereas the other coefficients in
the square brackets are, $(9N'\,|\mathrm{LHS})$ [which is necessary due to
Eq.~(\ref{RHS})] requires $(9\,|{N''}^2)$. Thus we need $N$ to be an
odd multiple of $27$, {\it i.e.} $N= 54 \cdot\mathbb{N}+27$. For such
$N$, all linearly allowed DGSs are consistent with the cubic anomaly
condition.

\item[$(iv)$] $\bsym{(9\,|N),\; N=\mathrm{even\!:}}$ From 
Table~\ref{lintab} we have in this case $n=N'$, $p=(1+3\ell_p)N''$ and
$m=(1+3s_m)\frac{N''}{2}$. The LHS then becomes
\beqn
\mathrm{LHS}&\!\!\!=\!\!\!&\Big[ -216 {\ell_p}^3 +108 {\ell_p}^2 
(-13+3s_m) + 18 \ell_p (-119 +18 s_m+9{s_m}^2) \notag \\ 
&\!\!\!\!\!\!& ~~+ (-1291 +585s_m-297
{s_m}^2+27{s_m}^3 ) \Big]\cdot \frac{N'}{2}  \cdot    \left( \frac{N''}{2}
\right)^{\!\!2}\!\!.
\eeqn
$1291$ is not a multiple of 9 (it is actually a prime), whereas the
remaining coefficients in square brackets are multiples of 9. Therefore the LHS is
not a multiple of $9\frac{N'}{2}$ in the case of $\neg\,(12\,|N)$,
respectively $9 N'$ in the case of $(12\,|N)$ [{\it cf.} 
Eq.~(\ref{RHS})], unless $(9\,|{N''}^2)$.  Thus the cubic anomaly
constraint requires $N\in54\cdot\mathbb{N}$ in this category. All
linearly allowed DGSs are possible for these values of $N$.
\end{itemize}
\end{enumerate}
Table~\ref{cubtab} in Sect.~\ref{cubic} summarises the results of this
appendix.

\end{appendix}

\bibliographystyle{hunsrt}

\bibliography{references}

\end{document}